\pdfoutput=1
\RequirePackage{fix-cm}
\documentclass{svjour3}
\smartqed
\usepackage{graphicx,xcolor}
\usepackage[numbers]{natbib}
\usepackage{amsmath,amssymb}
\usepackage{array}
\usepackage{booktabs}
\usepackage{siunitx}
\usepackage{subcaption}
\newcommand{\DSMC}{DSMC}
\newcommand{\SPARTA}{SPARTA}

\newcommand{\Kn}{Kn}
\newcommand{\Ma}{M}
\newcommand{\Rey}{Re}

\newcommand{\cpi}{c_{p,i}}

\journalname{Theoretical and Computational Fluid Dynamics}

\begin{document}

\title{Kinetic study of Reaction-Control-Jet Interference in Rarefied Hypersonic Crossflow}
\titlerunning{Kinetic study of reaction-control-jet interference}

\author{Alexis Theofilis \and Martin Grabe}
\authorrunning{A. Theofilis and M. Grabe}

\date{}

\institute{A. Theofilis \at
  Department of Aeronautics, Imperial College London, London, UK;
  work conducted at the German Aerospace Center (DLR), G\"ottingen, Germany\\
  \email{ait25@ic.ac.uk}\\
  ORCID: 0009-0008-6696-4642
  \and
  M. Grabe \at
  German Aerospace Center (DLR), Institute of Aerodynamics and Flow Technology,
  Spacecraft Department, G\"ottingen, Germany\\
  \email{martin.grabe@dlr.de}\\
  ORCID: 0000-0003-0361-2734
}

\maketitle

\begin{abstract}
A 1975 wind-tunnel study at the German Aerospace Center (DLR) of a helium jet blowing from a flat plate into a Mach 11 to 12 air crossflow at three freestream densities is revisited using two- and three-dimensional Direct Simulation Monte Carlo (DSMC) computations at Knudsen numbers 0.021 to 0.125. The planar slot used in two-dimensional model computations reproduces the shock topology, the sign and two-lobed structure of the interference pressure and its trend with rarefaction, and its plate-alone boundary layer is consistent with the three-dimensional one and with a similarity solution with slip and temperature jump. However, two-dimensional simulations overpredict the peak interference pressure about twelvefold at every condition and carry it farther upstream. A resolution sweep at the most rarefied condition converges the planar interference to within 1\% of its extrapolated limit. Three-dimensional DSMC reproduces the measured interference pressure at all three densities, within 25\% at the lobe peak and 3 mm in its position, although two- and three-dimensional simulations overpredict the plate-alone wall pressure by 50 to 75\%. Continuum breakdown around the jet persists even at the lowest Knudsen number examined. Fully three-dimensional kinetic simulations are underway to extend the present validation to a comparison of jet strength and forces over wider Knudsen number ranges, and to an examination of lateral relief as a reason for the discrepancy.
\end{abstract}

\section{Introduction}
\label{sec:intro}

Vehicles re-entering the Earth's atmosphere through the transitional rarefied regime are often steered with reaction control jets, small thrusters that fire straight into the oncoming hypersonic flow. The force the vehicle feels is not simply the thrust. The jet blocks and entrains the surrounding flow, which sets up a pressure field on the vehicle surface around the nozzle and adds an interference force to the commanded one \citep{spaid1973}. To date the most complete canonical experiment on this interaction is the work of Hefer \citep{hefer1975,hefer1976} whose 1975 study in the G\"ottingen hypersonic vacuum wind tunnel addressed interaction of jets in rarefied hypersonic crossflow with the oncoming laminar boundary layer on a flat plate. Hefer measured wall pressure, interference pressure, flow deflection, helium concentration and pitot profiles over a range of freestream densities and jet pressures. These results survived only as printed diagrams in an internal report and provided initial motivation for the present study. 
%
At the Knudsen numbers of Hefer's experiment ($\Kn_D=0.021$--$0.125$, based on the nozzle diameter) the flow is transitional, and Direct Simulation Monte Carlo (\DSMC{}) can be employed to provide reliable predictions. DSMC at the experimental conditions is also 
expensive. In the results reported in what follows, a converged three-dimensional simulation of a round jet on a finite plate was found to require $\sim3\times10^9$ particles and 23 hours on 7680 MPI ranks, for a total of $\sim1.8\times10^5$ core-hours. This raises the question whether a two-dimensional planar analogue of the actual experimental configuration, the computational cost of which is about 37 times lower than the full three-dimensional simulation (about $4.9\times10^{3}$ core-hours for the C1 case, against $1.8\times10^5$), can provide physically meaningful solutions, at least for some of the parameters reported. That is the second motivation of the present study.


The role of side jets is well recognized in the continuum supersonic and hypersonic literature. Spaid and Cassel \citep{spaid1973} established the experimental and theoretical framework for reaction-control interference, and their AGARDograph remains a primary reference for interference scaling. Wind-tunnel campaigns on re-entry and other vehicle configurations have since built up a large database of interference force coefficients and jet force amplification for specific vehicle geometries. Powrie, Ball and East \citep{powrie1993} compared two- and three-dimensional transverse jets in a hypersonic stream. Powrie's thesis \citep{powrie1996} used flat plates at a freestream Mach number of 6.7 under a laminar boundary layer, one with slot injection and one with axisymmetric injection. It found that the main characteristics of the upstream interaction are set by the size of the obstacle the jet presents to the oncoming flow, and that lateral venting from the separated region significantly affected the correlation of the slot-jet data. Dixon et al.\ \citep{dixon2005} simulated a transverse slot jet in two and three dimensions to assess the degree by which lateral venting changes the flow features and found that it strongly affects their extent. Mahesh \citep{mahesh2013interaction} reviews jets in crossflow, and Karagozian \citep{karagozian2014} shows that the upstream shear layer of the jet changes from convectively to absolutely unstable as the momentum flux ratio falls. Xu et al.\ \citep{Chenhao2014} reviewed the interference effects of reaction control systems on hypersonic vehicles, grouping the layouts into four categories (jets on surfaces of small and large curvature, jets on the nose, and jets near the wings) and summarising the state of the art for each. In the rarefied regime, \DSMC{} has been applied to a control-jet experiment at low density, e.g., in the study of Tartabini et al.\ \citep{tartabini1995}, but, to the best of our knowledge, there is no quantified comparison of planar and three-dimensional models against measurements.

A separate body of work concerns jets exhausting into rarefied or near-vacuum surroundings, where no crossflow is present. Dettleff \citep{Dettleff1991} reviewed plume flow and plume impingement in space technology, from the hypersonic continuum near field to the free-molecular far field and the forces and heat loads that an impinging plume imposes, and Yarygin et al.\ \citep{Yaryginetal2011} review three decades of work on the gas-dynamic and contaminating impact of thruster plumes on spacecraft and space stations. In this body of work the jet meets a surface in a quiescent or near-vacuum environment, so neither the blockage nor the entrainment mechanism that a crossflow produces can act. Kinetic simulation of hypersonic rarefied flow over a flat plate itself goes back to the earliest direct simulation work \citep{vogenitz1970,Vogenitzetal1971}, and \DSMC{} is now the established tool for such flows \citep{ivanov1998}, among them the SMILE system developed by Ivanov and co-workers at the Khristianovich Institute in Novosibirsk \citep{ivanov1998aiaa}. Jets issuing into a rarefied crossflow, however, have been treated much less. Tartabini et al.\ \citep{tartabini1993,tartabini1995} simulated an ESA experiment in the SR3 low-density tunnel at Meudon, in which a nitrogen jet is fired from one of two flat plates forming a corner in a Mach 20 stream. Their first report used a simplified jet at about a tenth of the experimental density and gave only qualitative insight \citep{tartabini1993}. Glass \citep{Glass2018} reports that the completed simulations gave jet-induced wall pressures above the measured values, and he attributes this to a grid too coarse for the much denser jet. The cost of resolving the dense jet led to the uncoupled CFD--\DSMC{} approach, in which a continuum jet plume computed by CFD is imposed as an inflow boundary on a \DSMC{} simulation of the interaction region. Glass and LeBeau \citep{glass1997} applied it to a sonic argon jet on a sharp-leading-edge flat plate in Mach 9.84 flow at a nozzle-diameter Knudsen number of 0.04, which is the configuration of an experiment in a low-density tunnel at Farnborough in the United Kingdom. Its data were unpublished at the time, so the comparison was between CFD and \DSMC{}. Liang et al.\ \citep{liang2014} simulated a sonic helium jet on a three-dimensional flat plate in Mach 13 nitrogen flow in a low-density tunnel at the China Aerodynamics Research and Development Center. They compared glow-discharge and oil-flow visualisations and the separation length for three jet pressures, and their calculated separation lengths were smaller than the measured ones by up to 10\%. Glass developed the technique further \citep{Glass2000} and, in a three-dimensional parametric study of a flat plate \citep{glass2004}, found that decreasing the freestream density increases the extent and magnitude of the normalised interference pressure, moves it upstream of the nozzle exit, and changes the sign of the jet-induced moment. Measurements for this configuration are scarce. Besides Hefer's, Staack et al.\ \citep{Staack2001} measured the temperature, pressure and velocity fields of a transverse jet on a flat plate in a rarefied hypersonic stream with planar laser-induced iodine fluorescence.

Kinetic studies of side jets on vehicle-like bodies are more recent. Gimelshein et al.\ \citep{gimelshein2002} studied a 3000-N-class thruster on the side of a small vehicle at altitudes of 80 and 100 km, including chemical reactions between the freestream and the plume gases. They computed the jet core with an axisymmetric calculation and the interaction with a three-dimensional \DSMC{} one. Kashkovsky et al.\ \citep{kashkovsky2014} used a multizone \DSMC{} approach for the control-thruster plume of the Intermediate eXperimental Vehicle and found that the interference changes the pitching moment by about 10\% below 120 km altitude. Liang et al.\ \citep{liang2014} extended their code to a slender blunt double cone, studying the effects of altitude, freestream speed, angle of attack and nozzle exit distribution on the flow field. Karpuzcu and Levin \citep{Karpuzcu2023} showed that \DSMC{} is needed to capture the multiscale shock and jet structure on a hypersonic cone even in a near-continuum freestream, because the strong gradients in the shocks and the jet expansion cause local continuum breakdown, and they found that the interaction is unsteady, with deviations on the order of 100\% in the drag force within the interaction region. They also report from earlier work that the interaction region is largest on a flat plate when flat-plate, wedge and cone geometries are compared at Mach 6.

Three gaps can be identified in the kinetic literature. First, to the best of our knowledge no kinetic simulation of a jet on a flat plate in hypersonic rarefied crossflow has been validated against a measured interference-pressure distribution, or against the interference force that distribution implies. The flat-plate simulations of Glass and co-workers could not be compared with measurements, Liang et al.\ compared flow visualisation and a single separation length, and the corner-flow \DSMC{} of Tartabini et al.\ overpredicted the measured pressure. Hefer's measurements, which resolve the interference pressure at three freestream densities, have not been reproduced numerically. Studies of vehicle-like bodies report flow fields and changes in the force and moment coefficients, but we know of none that resolves the interference into its blockage and entrainment contributions. Second, the planar representation of the jet that is common in simulation has not been quantified against the three-dimensional flow or against measurements, so it is unknown which interference quantities it can be trusted for. Third, the field has asked for exactly this kind of work. In his review of the uncoupled technique, Glass \citep{Glass2018} notes that jet interactions in the rarefied and transitional regimes with a hypersonic external flow have been studied much less than at either limit, and lists comparison with experiment and systematic studies on generic models such as a flat plate among the research still needed.

The present work makes four contributions. First, \DSMC{} is applied to the jet plume and the crossflow together, without a continuum--kinetic split, on a flat plate at the three freestream densities of Hefer's experiment, and the resulting interference pressure is compared quantitatively with his measurements. Second, the planar slot and the three-dimensional round jet are compared with each other and with those measurements, which shows what the planar idealisation can and cannot reproduce. Third, the spanwise velocity field of the three-dimensional reference points to lateral relief as the likely mechanism behind the planar overprediction and measures how much flow it diverts. Fourth, the plate-alone boundary layer, which is the base state of the interaction, is checked against a similarity solution with slip and temperature jump and against earlier kinetic work. The broader three-dimensional validation against Hefer's flow deflection, helium concentration and pitot data, the integration of the interference into forces and moments, and the sweeps in rarefaction and jet strength belong to the companion study and are outside the scope of this paper.

We examine the planar slot as a \DSMC{} model of the interaction, at all three freestream conditions of Hefer \citep{hefer1975,hefer1976}, against his measurements and against three-dimensional reference kinetic simulations also performed here. Both the two- and the three-dimensional simulations were performed with the \SPARTA{} solver \citep{sparta}. The three-dimensional simulation, on the full asymmetric plate, is the reference. It reproduces the measured interference pressure, which shows that the planar discrepancy comes from the geometry, and its spanwise velocity field points to the mechanism. The full validation is reported in a companion study \citep{theofilis2026companion}, and preliminary results of this work were presented at the 9th International Conference on Jets, Wakes and Separated Flows \citep{theofilis2026icjwsf} and at the 79th Annual Meeting of the APS Division of Fluid Dynamics \citep{theofilis2026dfd}. The planar slot gets the shock topology, the plate-alone boundary layer, the sign and two-lobed structure of the interference pressure, and its trend with rarefaction right. It overpredicts the peak interference pressure by about a factor of 12 and carries it much farther upstream. We propose that the main cause is the missing lateral relief. A slot is an unbroken spanwise obstacle, so the crossflow can only pass over it. In the three-dimensional runs the jet turns aside 33 to 40\% of the flow aimed at a strip two nozzle diameters wide, at $\SI{5}{mm}$ above the wall, a fraction that is zero for a slot by construction.

Section~\ref{sec:problem} sets out the problem and Section~\ref{sec:method} the method and test cases. Section~\ref{sec:2d} reports the planar simulations, jet-off first and then jet-on. The jet-off results cover the wall pressure and the boundary layer, which we compare with a similarity solution and with the work of Klothakis et al. Section~\ref{sec:3dref} presents the three-dimensional simulations in the same order and compares each result with its planar counterpart, ending with the interference pressure. Section~\ref{sec:2d_mechanism} examines the origin of the discrepancy and Section~\ref{sec:conclusions} concludes.

\section{Problem and background}
\label{sec:problem}

\begin{figure}[h]
  \centering
  \includegraphics[width=0.6\linewidth]{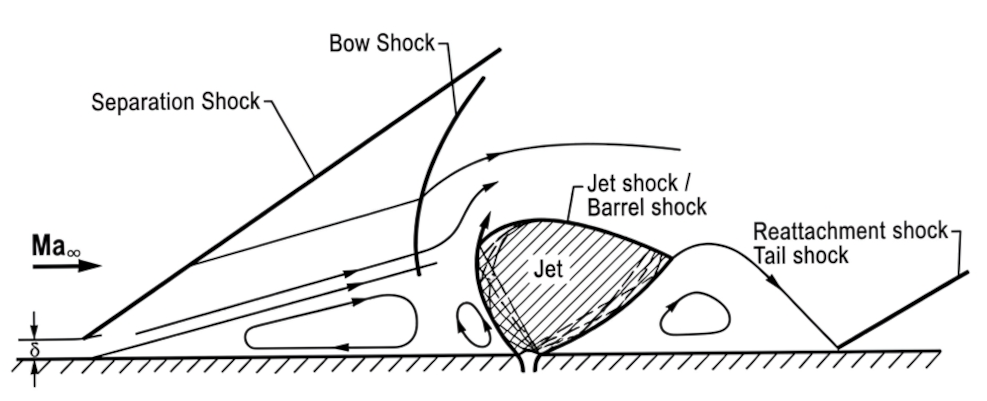}
  \caption{Flow around a reaction control jet in hypersonic crossflow, redrawn and translated from the original diagram \cite{hefer1975}.}
  \label{fig:RCSfig}
\end{figure}

\begin{figure}[htbp]
  \centering
  \includegraphics[width=0.75\linewidth]{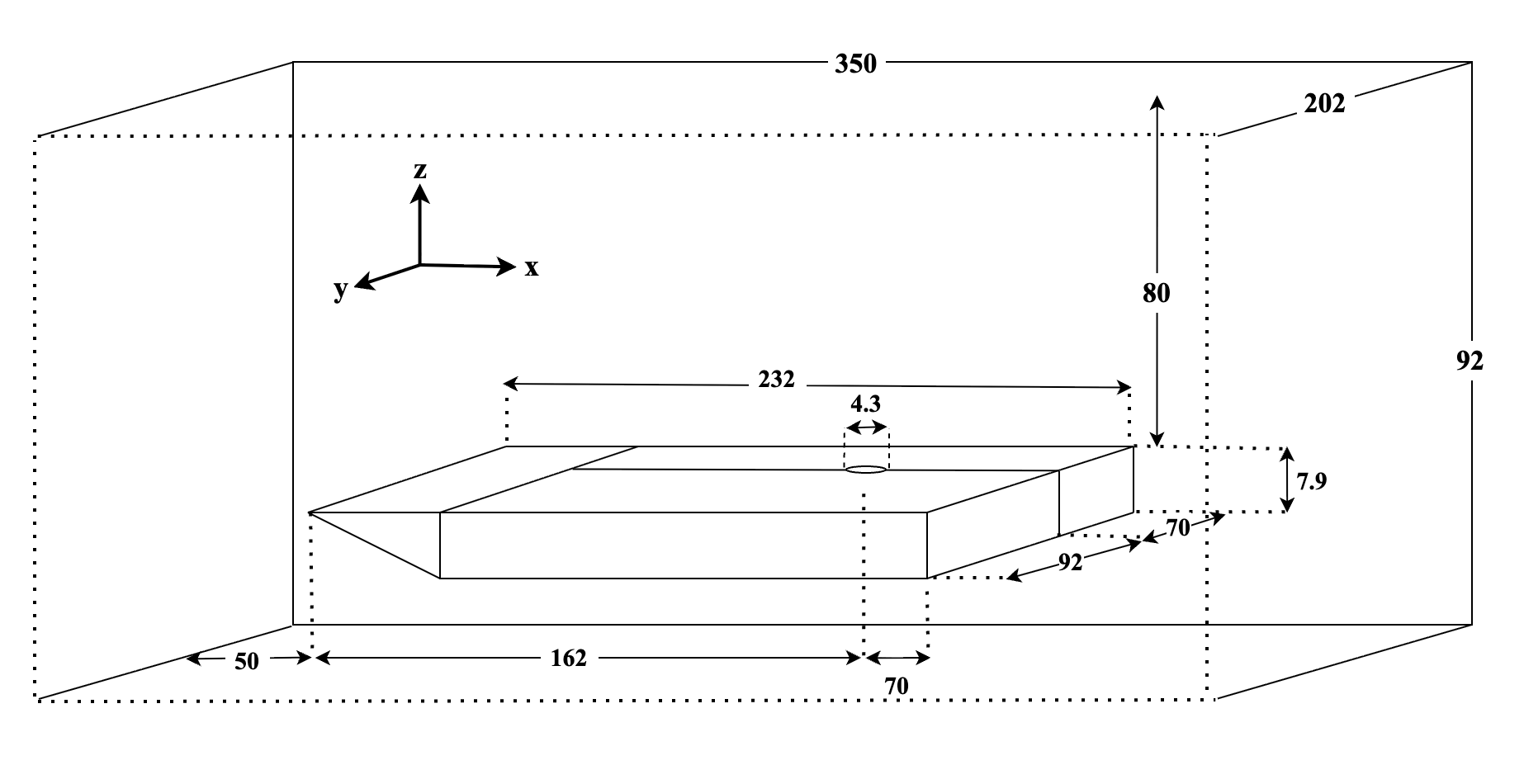}
  \caption{Sketch of the full asymmetric three-dimensional domain, reproducing the geometry of Hefer \citep{hefer1975}. All dimensions are in millimetres and not to scale. There is no symmetry plane, and all faces except the downstream one are freestream boundaries.}
  \label{fig:dom_asym}
\end{figure}

Figure~\ref{fig:RCSfig} sketches the configuration, a reaction control jet issuing normal to a flat plate into a hypersonic crossflow. Figure~\ref{fig:dom_asym} shows the geometry and domain of the three-dimensional model. Coordinates follow Hefer and DLR practice, with $x$ streamwise, $y$ spanwise and $z$ wall-normal. The planar runs have only $x$ and $z$. The streamwise origin changes between figures, so we state it each time. The coordinate $\bar x$ is measured from the leading edge, $x$ from the nozzle centre in the interference-pressure and profile plots, and $x$ from the inflow face of the domain in the flow-field plots of the plane through the nozzle ($y=0$) (Figures~\ref{fig:2d_streamlines} and~\ref{fig:field_sym_jeton}). Because we use kinetic theory throughout, the regions where the continuum assumption holds and those where it breaks down are computed in one framework.

\subsection{Jet-in-crossflow at rarefied conditions}
\label{sec:jicf}

Jet-in-crossflow is a classic continuum problem. Mahesh \citep{mahesh2013interaction} reviews it, and Karagozian \citep{karagozian2014} examines the stability of its shear layer. The flow is dominated by vortices, including a counter-rotating vortex pair, a horseshoe vortex ahead of the orifice and ring vortices shed from the shear layer. At supersonic and hypersonic continuum conditions the same structures persist alongside a bow shock, a barrel shock and a Mach disk in the underexpanded plume, and an upstream separation region \citep{powrie1996}. Hefer's conditions are transitional ($\Kn_D=0.021$--$0.125$, Table~\ref{tab:cases}), so we should not expect the continuum structures to vanish altogether. But a vortex is a continuum object at the scale of its own size. The small separation region near the nozzle can therefore have a local Knudsen number, based on its own dimension, that is far from small even when the bulk flow is only mildly transitional. Hefer notes that ahead of the jet one cannot speak of a vortex in the classical sense, since the mean free path is large compared with that region and continuum assumptions do not appear admissible \citep{hefer1975}. Whether a closed streamline marks a vortex in the continuum sense has to be judged against the local breakdown parameter $\Kn_\text{GLL}$ (Section~\ref{sec:breakdown}), not the bulk Knudsen number.

\subsection{Blockage and entrainment}
\label{sec:mechanisms}

Hefer identified two ways in which the jet changes the wall pressure. In \emph{blockage}, the jet acts as an obstacle to the crossflow. A bow wave forms ahead of the nozzle and produces an upstream overpressure, which peaks a few nozzle diameters ahead of the exit and decays roughly exponentially with distance. In \emph{entrainment}, the free shear layer at the jet boundary sweeps ambient gas downstream. A sink-like inflow from the surroundings has to replace that gas, and this produces the pressure deficit downstream of the nozzle. Both mechanisms, and their signature in the wall pressure, hold across the whole range of conditions examined here, and the discussion below follows them.

\subsection{Interference pressure coefficient}
\label{sec:metrics}

Our main quantity is the interference pressure coefficient
\begin{equation}
  \cpi = \frac{p - p_P}{q_\infty},
  \label{eq:cpi}
\end{equation}
where $p$ is the wall pressure with the jet active, $p_P$ is the wall pressure at the same location with the jet off, and $q_\infty = \tfrac{1}{2}\rho_\infty u_\infty^2$ is the freestream dynamic pressure. It isolates the jet's effect from the boundary layer on the plate, however nonlinear the base flow is. Integrated over the plate it gives the interference force, which the companion study \citep{theofilis2026companion} integrates and compares with the jet's thrust (Section~\ref{sec:conclusions}). A planar simulation gives only a force per unit span, so we compare planar and three-dimensional results through $\cpi$ itself. We characterise the jet strength by the momentum flux ratio $J=\rho_j u_j^2/(\rho_\infty u_\infty^2)$.

\section{Numerical method and test cases}
\label{sec:method}

\subsection{DSMC and SPARTA}
\label{sec:method_dsmc}

We use \DSMC{} \citep{bird1994} as implemented in the open-source code \SPARTA{} \citep{sparta}. The gas is represented by computational particles, each standing for $F_\text{num}$ real molecules, and their motion and collisions are decoupled over a timestep. This is valid as long as a particle crosses only a small fraction of a cell and collides at most once per step, which requires $\Delta t<\tau_\text{coll}$ and a cell size below the mean free path $\lambda$. Appendix~\ref{app:verification} verifies both. Collisions use the no-time-counter scheme with the variable soft sphere model, rotational exchange follows Borgnakke--Larsen, and vibrational excitation and chemistry are switched off (Appendix~\ref{app:vss}). The temperatures justify this. The freestream total temperature, which bounds the static temperature of an adiabatic flow, is about \SI{800}{K} at C1 and \SI{600}{K} at C2 and C3, and the highest translational temperature in the planar C1 field is \SI{570}{K}, in a few cells at the leading edge, against \SI{445}{K} at the 99.9th percentile. At \SI{570}{K} the equilibrium vibrational energy per mode is about 8\% of the translational energy for O$_2$ and 2\% for N$_2$, and it is lower over almost all of the flow, so freezing it neglects an energy sink of a few percent at most. Dissociation and ionisation need temperatures several times higher. \DSMC{} makes no closure assumption about the velocity distribution function, so a single domain can hold the free-molecular leading edge, the strongly interacting mid-plate and the near-continuum jet core without a hybrid continuum--kinetic treatment \citep{Glass2000}. \SPARTA{} uses a hierarchical Cartesian grid with surface cut cells, adaptive mesh refinement and recursive-coordinate-bisection load balancing.

\subsection{Test cases}
\label{sec:method_cases}

Three freestream conditions, C1 to C3, reproduce the three main cases Hefer ran (Table~\ref{tab:cases}). C1 is the most rarefied and C3 the most continuum. The freestream number density rises by a factor of 5.5 from C1 to C3, and the nozzle-diameter Knudsen number $\Kn_D=\lambda_\infty/D$ falls from 0.125 to 0.021. As in Hefer's own tunnel campaign, the stagnation conditions change along with the density. The jet is the same in all three cases (helium at $p_{s0} = \SI{50}{Torr}$, expanded through Hefer's $F_{sD}/F_s^\ast = 16$ Laval nozzle to $\Ma_D = 6.02$), so the relative jet strength falls from C1 to C3.

\begin{table}[htbp]
  \centering
  \caption{Freestream conditions as prescribed in the \SPARTA{} decks. $\lambda_\infty$ is the VSS mixture mean free path and $\Kn_D = \lambda_\infty/D$. Air is 78\% N$_2$ / 22\% O$_2$ by number fraction. The jet is the same in all three cases, helium at $p_{s0}=\SI{50}{Torr}$ and $T_{s0}=\SI{293}{K}$, expanding to $\Ma_D=6.02$.}
  \label{tab:cases}
  \small
  \setlength{\tabcolsep}{4pt}
  \begin{tabular}{@{}lcccccccccc@{}}
    \toprule
    & $p_0$ & $T_0$ & $\Ma_\infty$ & $n_\infty$ & $u_\infty$ & $T_\infty$
      & $\lambda_\infty$ & $\Kn_D$ & $q_\infty$ & $n_D/n_\infty$ \\
    & [ata] & [K] & & [m$^{-3}$] & [m/s] & [K] & [mm] & & [Pa] & \\
    \midrule
    \textbf{C1} (most rarefied)  & 0.6 & 800 & 11.3 & \num{1.490e21} & 1241.0 & 30.1 & 0.536 & 0.125 & 55.0  & 14.6 \\
    \textbf{C2} (intermediate)   & 1.0 & 600 & 11.7 & \num{2.840e21} & 1076.7 & 21.1 & 0.258 & 0.060 & 79.0  & 7.6  \\
    \textbf{C3} (most continuum) & 3.0 & 600 & 11.8 & \num{8.140e21} & 1078.2 & 20.8 & 0.090 & 0.021 & 227.0 & 2.7  \\
    \bottomrule
  \end{tabular}
\end{table}

\subsection{Three-dimensional reference configuration}
\label{sec:method_domains}

The reference reproduces Hefer's Model~I. This is a flat plate \SI{232}{mm} long and \SI{7.90}{mm} thick, with a \SI{7.5}{\degree} bevelled leading edge and the nozzle \SI{162}{mm} behind the leading edge. We model the nozzle as a spherical cap of half-angle \SI{8.5}{\degree} with an exit diameter of \SI{4.3}{mm}, four times Hefer's geometric throat diameter of \SI{1.075}{mm} (from his area ratio of 16). Hefer's drawing of the nozzle shows an exit bore of \SI{4.0}{mm}. The jet is emitted along the local normal of the cap, so it diverges as a conical source of \SI{8.5}{\degree} half-angle. The domain is the full asymmetric one (Figure~\ref{fig:dom_asym}). The plate has its true span and the nozzle sits off-centre, as on Hefer's plate (\SI{70}{mm} from one edge, \SI{92}{mm} from the other), because a mirrored half-domain would impose a symmetric plate. All six faces are outflow boundaries, and the freestream is emitted from all of them except the downstream one. The plate surface is a fully diffuse isothermal wall at $T_w = \SI{293}{K}$, standing in for Hefer's water-cooled copper plate, and the jet enters through a surface emission boundary on the nozzle cap at the exit state of Table~\ref{tab:cases}.

\subsection{Mesh, particle weighting and verification}
\label{sec:method_mesh}

Each run starts on a uniform \SI{2}{mm} base grid and passes through equilibration, thermal marking (two refinement passes keyed to temperature and pressure), a cell-width cascade to \SI{1.5}{mm}, and a final settling and production phase with $\Delta t = \SI{1e-8}{s}$ and cell-volume particle weighting. We anchor the particle weight to 1.5 particles per cell in the finest wall-adjacent shock cell, which holds the asymmetric C1 jet-on case to about $3\times10^9$ particles and 23\,h on 7680 ranks. The results rest on three resolution conditions (timestep against collision time, cell size against mean free path, and sampling time against flow-through time) and on statistical convergence, summarised in Appendix~\ref{app:verification}. \label{sec:method_verification}

\section{Planar-slot simulations}
\label{sec:2d}

This section describes the planar configuration and its results. We start with the plate-alone (jet-off) flow, which is the base state the jet acts on, first its wall pressure and then its boundary layer, and finish with the jet-on flow field. Section~\ref{sec:3dref} then compares these results with the three-dimensional simulations, in the same order.

\subsection{Configuration}
\label{sec:2d_config}

The planar campaign uses the same three freestream conditions (Table~\ref{tab:cases}) and the same jet stagnation state and exit Mach number as the three-dimensional reference ($p_{s0}=\SI{50}{Torr}$, $T_{s0}=\SI{293}{K}$, $\Ma_D=6.02$, which gives an exit velocity of $\SI{1676}{m/s}$ and a temperature of $\SI{22.4}{K}$). The round nozzle is replaced by a planar slot $\SI{4.0}{mm}$ wide (the three-dimensional exit diameter is $\SI{4.3}{mm}$) at the same station, $\SI{162}{mm}$ from the leading edge. We set the width from the exit geometry of the nozzle and the isentropic area-ratio relation, and we do not model the nozzle interior. The slot is an analogue of the round nozzle, not a reduction of it. A slot and an orifice of the same width do not issue the same mass flow, and the planar geometry removes lateral relief entirely. All other physical models are those of Section~\ref{sec:method}. In both the planar and the three-dimensional domains the wall extends beyond the physical plate length of $\SI{232}{mm}$, so we limit comparisons to $\bar x\le\SI{23.2}{cm}$ and $x\le\SI{60}{mm}$ from the nozzle, which lie on the physical plate in both.

The two jets differ in one respect. The three-dimensional campaign prescribes an exit number density of $\SI{2.17e22}{m^{-3}}$. This is the ideal isentropic value of $\SI{3.48e22}{m^{-3}}$ reduced by Hefer's measured discharge coefficient ($\zeta=0.62$ at $p_{s0}=\SI{50}{Torr}$), so that the simulated jet carries the mass flow of his real nozzle. The planar decks use the ideal value because no nozzle is modelled. The planar jet is therefore 1.6 times denser and carries 1.6 times the momentum flux per unit area. The momentum flux ratio $J$ is 5.9, 4.1 and 1.4 for C1 to C3 in two dimensions, against 3.7, 2.6 and 0.9 in three. If the interference pressure scales roughly linearly with jet strength, as the companion study \citep{theofilis2026companion} finds at transitional conditions, this explains a factor of about 1.6 of the 11 to 15 times difference in the peak (Section~\ref{sec:2d_interference}) and leaves a factor of about 7 to 9 for the geometry. A planar run at the three-dimensional exit density would remove this difference. The planar C1 jet-on run took $4.9\times10^{3}$ core-hours, a factor of 37 less than its three-dimensional counterpart (Appendix~\ref{app:verification}).

The domain is a rectangle $\SI{442.9}{mm}$ long and $\SI{250}{mm}$ high, periodic in the third direction. Its base grid of $688\times400$ cells is refined through four levels to a finest cell size of $L4=\SI{80.5}{\micro m}$, and the refinement follows the solution, as in the reference. All four faces are outflow boundaries, so molecules that reach them leave the domain, and the freestream enters by emission from the left face, the top face and the part of the bottom face ahead of the plate ($0\le x\le\SI{50}{mm}$). The plate is a zero-thickness line lying flush with the lower boundary from $x=\SI{50}{mm}$ to $\SI{442.9}{mm}$ ($\SI{432.6}{mm}$ in the C3 jet-on run), with a fully diffuse isothermal wall at $\SI{293}{K}$. Nothing downstream of $x=\SI{282}{mm}$, where Hefer's plate ends, enters the comparisons. The slot is a $\SI{4.0}{mm}$ segment of that line at $x=\SI{212}{mm}$, from which the jet gas is emitted at the exit state given above as a parallel stream normal to the wall. This differs from the three-dimensional cap, whose jet diverges with a half-angle of $\SI{8.5}{\degree}$. C1 and C2 use a staggered particle-weighting scheme. The grid is marked cheaply at a high $F_\text{num}$, which is then reduced by a factor of 5.46 before production sampling. C3 goes straight from marking to a three-transit settle and production, because its higher density already gives enough particles per cell. The jet-off runs of all three cases are sampled at the marking weight, which leaves about 0.8 simulated particles in a finest freestream cell and sets the statistical noise of every plate-alone quantity below (Appendix~\ref{app:verification}). The resolution audit of C3 puts its post-shock region at $L4/\lambda \approx 4.2$, against the usual target of $\lambda/3$. We accepted this compromise, as the reference did in its shock interior (Appendix~\ref{app:verification}). Table~\ref{tab:2d_repro} lists the parameters needed to reproduce the runs.

\begin{table}[htbp]
  \centering
  \caption{Planar campaign parameters, C1 to C3. Transit time is $L_x/u_\infty$, and settle and production step counts are quoted at the production timestep $\Delta t = \SI{1e-8}{s}$. The jet cases of C1 and C2 use a develop, re-mark and re-settle sequence before production (described in the text). C3 proceeds directly from marking to a single settle. The jet-off runs of all three cases are sampled at the marking weight, and the PPC row refers to the jet-on runs.}
  \label{tab:2d_repro}
  \footnotesize
  \setlength{\tabcolsep}{3pt}
  \resizebox{\textwidth}{!}{%
  \begin{tabular}{@{}lccc@{}}
    \toprule
    & C1 & C2 & C3 \\
    \midrule
    $F_\text{num}$, marking       & \num{1.20e13} & \num{2.29e13} & \num{6.55e13} \\
    $F_\text{num}$, production    & \num{2.20e12} & \num{4.19e12} & \num{6.55e13} (unchanged) \\
    Transit time                  & \SI{354.6}{\micro s} & \SI{411.3}{\micro s} & \SI{410.8}{\micro s} \\
    Jet develop steps (transits)  & 177{,}000 (5) & 205{,}000 (5) & none \\
    Jet re-settle steps (transits)& 107{,}000 (3) & 123{,}000 (3) & 122{,}500 (3), direct \\
    Jet production (2 windows)    & $2\times\num{66400}$ & $2\times\num{76600}$ & $2\times\num{76400}$ \\
    Freestream settle (transits)  & \num{50000} (1.40) & \num{57500} (1.40) & \num{57500} (1.40) \\
    Freestream production         & $2\times\num{66400}$ & $2\times\num{76600}$ & $2\times\num{76400}$ \\
    Phase-1 pressure threshold    & \SI{0.93}{Pa} ($1.5p_\infty$) & \SI{1.24}{Pa} & \SI{3.51}{Pa} \\
    PPC at L4 (fs/post-shock/He)  & $\sim$4.4/$\sim$26/n.a. & $\sim$4.4/$\sim$26/n.a. & $\sim$0.8/$\sim$4.8/$\sim$18.6 \\
    \bottomrule
  \end{tabular}%
  }
\end{table}

\subsection{Plate-alone wall pressure}
\label{sec:2d_wallpressure}

\begin{figure}[htbp]
  \centering
  \includegraphics[width=\linewidth]{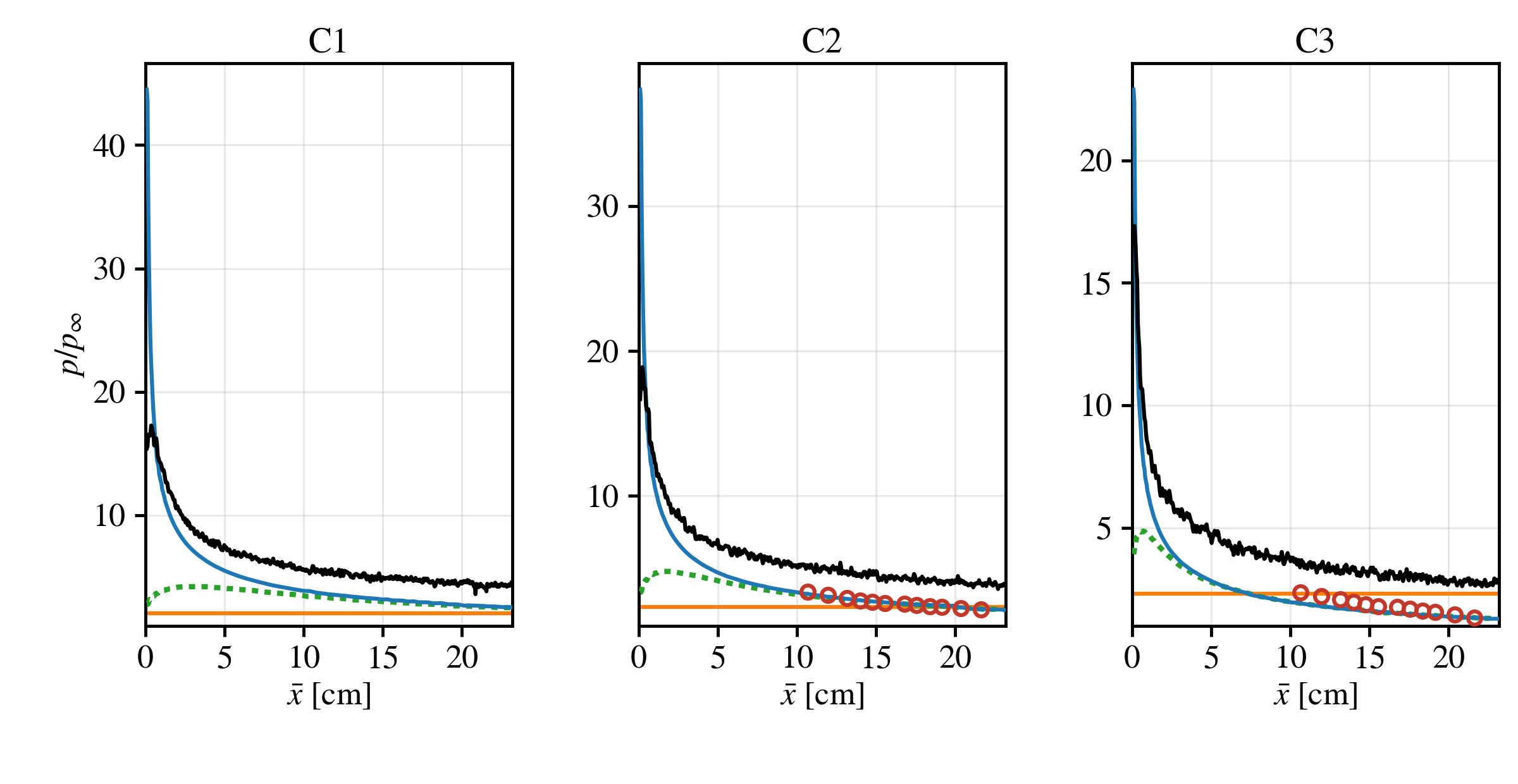}
  \caption{Planar plate-alone wall pressure for C1, C2 and C3 (left to right). The black line shows the \DSMC{} result and the red circles Hefer's measurements (C2 and C3). The blue, orange and green dotted lines are the Cheng (continuum), Schaaf--Chambre (free-molecule) and Kienappel (blend) correlations.}
  \label{fig:2d_wallpressure}
\end{figure}

Figure~\ref{fig:2d_wallpressure} shows the planar plate-alone wall pressure against three correlations Hefer used and against his measurements, which exist for C2 and C3 only. The correlations are Cheng's for the continuum limit, Schaaf and Chambre's for the free-molecule limit, and Kienappel's blend of the two \citep{hefer1976}. The pressure lies above all three over most of the plate. It is also well above Hefer's measurements. At his twelve stations ($\bar x=10.6$ to $\SI{21.6}{cm}$) it is on average 1.70 times his value for C2 (range 1.54 to 1.84) and 1.75 times for C3 (1.49 to 2.00), and the ratio rises along the plate, from 1.57 to 1.80 for C2 and from 1.55 to 1.95 for C3 between $\bar x=10.7$ and $\SI{21.6}{cm}$. Hefer attributes a similar deviation from theory in his own data to the non-parallel flow of the tunnel nozzle \citep{hefer1976}. A source-flow estimate shows that this cannot be the whole explanation. It would require the freestream static pressure to fall by 24\% (C2) and 38\% (C3) over the plate, which is a lot for a flow Hefer calls weakly expanding, and the \DSMC{} would still be 1.36 and 1.21 times his values. Rarefaction corrections, which he did not apply and which would raise his pressures, are another candidate. The cause of the offset remains open.

\subsection{Plate-alone boundary layer}
\label{sec:2d_bl}

\begin{figure}[htbp]
  \centering
  \includegraphics[width=\linewidth]{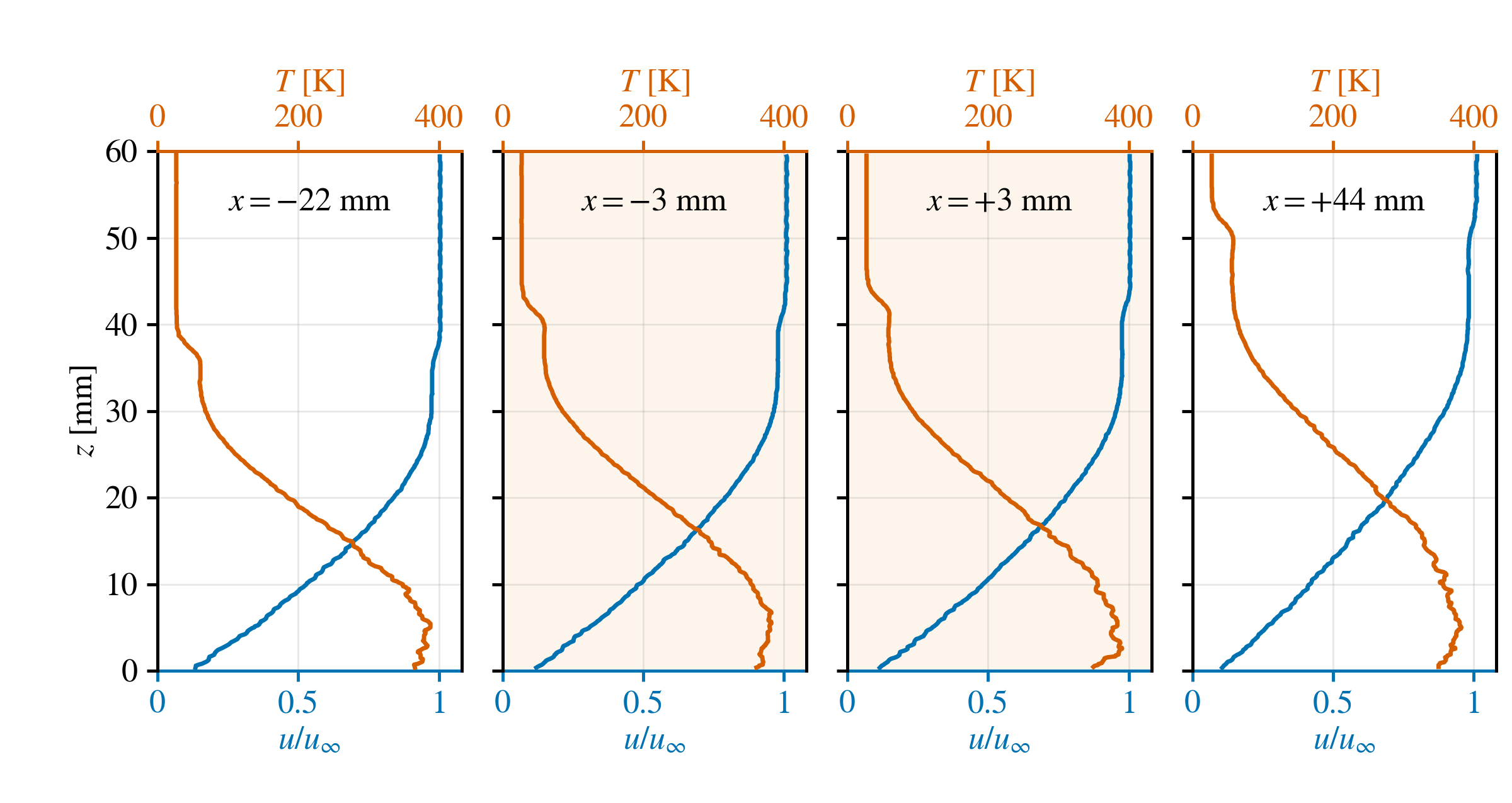}
  \caption{Wall-normal profiles of the streamwise velocity (blue, bottom axis) and the temperature (orange, top axis) at four of the eight streamwise stations, case C1, jet-off. The coordinate $x$ is measured from the nozzle position, and the shaded plates ($x=-3$ and $+\SI{3}{mm}$) are closest to the nozzle position.}
  \label{fig:2d_bl_profiles}
\end{figure}

The boundary layer of the same jet-off simulations completes the check of the base state the jet acts on. We extracted wall-normal profiles of temperature and streamwise velocity at eight stations from $x=-22$ to $+\SI{44}{mm}$ relative to the nozzle position, for all three cases. Figure~\ref{fig:2d_bl_profiles} shows case C1 at four of them, the farthest upstream and downstream and the two adjacent to the nozzle. The intermediate stations vary smoothly between these. At every station both quantities go smoothly from the wall to the freestream. The velocity $u$ rises from near zero at the wall to $u_\infty$ over a layer that thickens gradually downstream, and $T$ falls from a near-wall maximum of order $\SI{400}{K}$ through a shock-heated shoulder to the freestream value. The only visible scatter, in the first few millimetres above the wall, looks like \DSMC{} sampling noise. C3, with its higher freestream density and better particle statistics, has the same shape with visibly less noise and a thinner layer. The jet-off runs therefore give a clean base state for the interference coefficient (Eq.~\ref{eq:cpi}).

\begin{figure}[htbp]
  \centering
  \includegraphics[width=\linewidth]{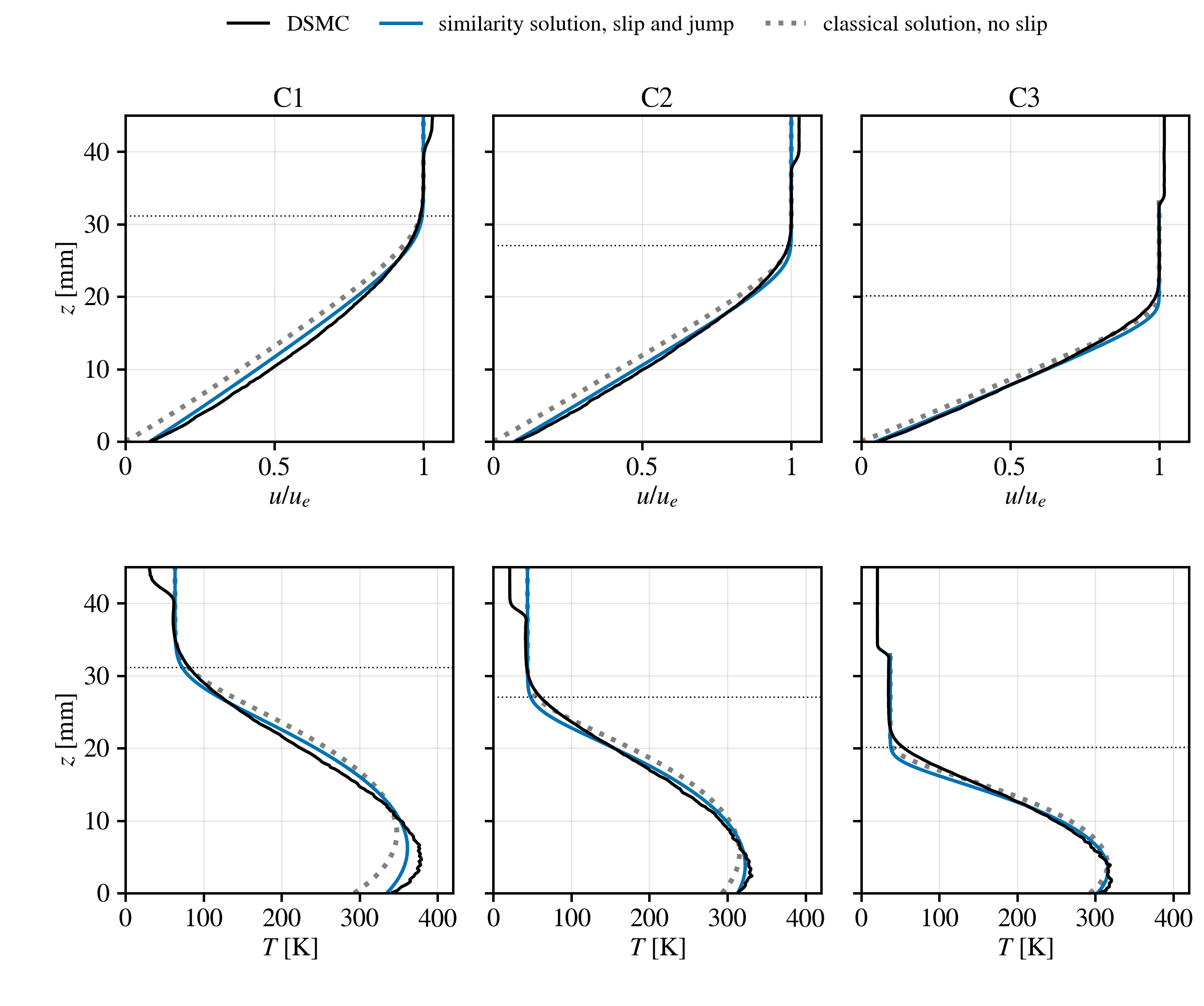}
  \caption{Planar jet-off profiles of the streamwise velocity (upper row) and temperature (lower row) at the nozzle station, $\SI{162}{mm}$ from the leading edge, for C1, C2 and C3 (left to right). The black lines show the \DSMC{} profiles and the blue lines the similarity solution with the slip velocity and temperature jump of the \DSMC{} field. The grey dotted lines show the classical no-slip solution with $T_w=\SI{293}{K}$ on the same edge state. The thin dotted horizontal line marks $\delta_{99}$ of the \DSMC{} profile. Above the layer the \DSMC{} profile continues through the leading-edge shock into the freestream.}
  \label{fig:cbl_noz}
\end{figure}

Next we compare the layer with a compressible laminar similarity solution. For each case the solution is built on the \DSMC{} edge state, taken at the thermal edge of the layer, with the slip velocity and temperature jump extracted from the \DSMC{} field imposed at the wall. As a reference without slip we use the classical no-slip solution at $T_w=\SI{293}{K}$ on the same edge state. The equations use a constant Prandtl number of 0.72, $\gamma=1.4$ and a power-law viscosity $\mu\propto T^{0.74}$, with the reference value from Wilke's mixture rule applied to the VSS viscosities of N$_2$ and O$_2$. Because the edge state, the slip and the jump come from the \DSMC{} field, the comparison checks that the \DSMC{} profile is consistent with continuum boundary-layer theory given these inputs, and it does not validate the \DSMC{} independently. Figure~\ref{fig:cbl_noz} compares the profiles at the nozzle station, and Table~\ref{tab:bl_stations} compares the two at three stations per case.

\begin{table}[htbp]
  \centering
  \caption{Similarity solution with the slip and temperature jump taken from the \DSMC{} field, against the \DSMC{} profile, at three stations per case. The coordinate $x$ is the distance from the leading edge, and the nozzle is at $\SI{162}{mm}$. The rms velocity differences over the layer are normalised by the edge velocity. The classical solution has no slip and $T_w=\SI{293}{K}$, on the same edge state.}
  \label{tab:bl_stations}
  \small
  \setlength{\tabcolsep}{4pt}
  \begin{tabular}{@{}lrrrrrrrr@{}}
    \toprule
    & & & & \multicolumn{2}{c}{$\delta_{99}$ [mm]} & & \multicolumn{2}{c}{rms $u/u_e$ [\%]} \\
    \cmidrule(lr){5-6}\cmidrule(lr){8-9}
    Case & $x$ [mm] & $\sqrt{\Rey_{x,e}}$ & $u_\text{slip}/u_e$ [\%] & \DSMC{} & similarity & ratio & no slip & slip + jump \\
    \midrule
    C1 & 39 & 24.1 & 15.2 & 9.2 & 14.5 & 1.57 & 28.3 & 19.5 \\
     & 162 & 79.2 & 8.2 & 31.2 & 29.9 & 0.96 & 7.5 & 3.0 \\
     & 231 & 92.9 & 7.1 & 40.1 & 38.3 & 0.95 & 6.0 & 2.3 \\
    \addlinespace
    C2 & 39 & 49.6 & 12.9 & 9.4 & 10.3 & 1.10 & 16.7 & 8.6 \\
     & 162 & 111.8 & 7.1 & 27.1 & 25.4 & 0.94 & 6.3 & 2.0 \\
     & 231 & 129.3 & 5.7 & 34.9 & 33.3 & 0.95 & 5.4 & 2.3 \\
    \addlinespace
    C3 & 39 & 95.1 & 8.6 & 7.3 & 6.8 & 0.94 & 7.8 & 2.6 \\
     & 162 & 178.7 & 5.0 & 20.1 & 18.2 & 0.90 & 3.9 & 2.1 \\
     & 231 & 208.4 & 3.4 & 25.7 & 23.5 & 0.91 & 3.3 & 2.1 \\
    \bottomrule
  \end{tabular}
\end{table}

At and downstream of the nozzle the similarity solution describes the layer well. At the nozzle station the rms velocity difference over the layer is 3.0, 2.0 and 2.1\% of the edge velocity for C1 to C3, against 7.5, 6.3 and 3.9\% for the classical solution. The similarity layer is slightly thinner than the \DSMC{} layer, with $\delta_{99}$ ratios of 0.96, 0.94 and 0.90, and the wall slip is 8.2, 7.1 and 5.0\% of the edge velocity. Figure~\ref{fig:cbl_noz} shows that the solution with slip and jump follows both the velocity and the temperature profile through the layer, while the classical solution departs from them near the wall. It also reproduces the temperature overshoot above the wall value. Its peak lies 4.4, 2.1 and 0.9\% below the \DSMC{} value for C1 to C3 (8.0, 4.3 and 1.1\% for the classical solution) and about 1 to 2 mm higher. The agreement is the same $\SI{69}{mm}$ farther downstream ($\delta_{99}$ ratios 0.95, 0.95 and 0.91, velocity differences 2.3, 2.3 and 2.1\%).

Upstream, the agreement depends on the Reynolds number. At $x=\SI{39}{mm}$ from the leading edge the similarity solution is markedly thicker than the \DSMC{} layer in C1 ($\delta_{99}$ ratio 1.57, velocity difference 19.5\%) and thicker in C2 (1.10, 8.6\%), while C3, with $\sqrt{\Rey_{x,e}}=95$, stays close (0.94, 2.6\%). This is expected where the viscous layer merges with the shock. The edge pressure is not constant along the plate, and the outer flow between the layer and the shock is graded rather than uniform. A zero-pressure-gradient similarity solution cannot represent either, and the remaining offset in the outer layer is consistent with that, not with the wall condition.

Klothakis et al.\ \citep{Klothakis2022} computed the same kind of layer with \SPARTA{} (VHS collisions, diffuse walls) for air, argon and nitrogen over a plate with a $\SI{0.5}{mm}$ nose radius, at freestream Mach numbers from 4.7 to 9.4. They compared it with a similarity solution that carries slip and temperature-jump boundary conditions. Their profile comparison covers the two adiabatic cases, air and argon at $x=\SI{0.7}{m}$. They report that the wall values from their slip and temperature-jump relations agree with those computed in the \DSMC{} for all three cases, and an earlier conference paper gives isothermal profile comparisons \citep{klothakis2021aiaa}. Their Case~3, isothermal nitrogen, is the closest to the present flows, and Table~\ref{tab:bl_klothakis} sets it against the nozzle station of C1 to C3.

\begin{table}[htbp]
  \centering
  \caption{The nozzle station of C1 to C3 against Case~3 of Klothakis et al.\ \citep{Klothakis2022} (isothermal nitrogen). The quantities $\sqrt{\Rey_{x,e}}$ and $\Ma_e$ are edge-based. The present edge is the thermal edge, whereas the edge definition of Klothakis et al.\ is not stated. $T_{g,w}$ is the gas temperature at the wall.}
  \label{tab:bl_klothakis}
  \small
  \begin{tabular}{@{}lcccc@{}}
    \toprule
     & C1 & C2 & C3 & Klothakis et al., Case~3 \\
    \midrule
    Gas & air & air & air & N$_2$ \\
    $\Ma_\infty$ & 11.3 & 11.7 & 11.8 & 9.39 \\
    $T_w$ [K] & 293 & 293 & 293 & 245 \\
    $T_w/T_0$ & 0.37 & 0.49 & 0.49 & 0.054 \\
    Station, distance from leading edge [mm] & 162 & 162 & 162 & 40 \\
    $\sqrt{\Rey_{x,e}}$ & 79.2 & 111.8 & 178.7 & 78.1 \\
    $\Ma_e$ & 7.52 & 7.90 & 8.65 & 6.99 \\
    $T_e/T_\infty$ & 2.11 & 2.08 & 1.79 & 1.72 \\
    $p_e/p_\infty$ & 5.61 & 5.01 & 3.38 & 3.39 \\
    $u_\text{slip}/u_e$ [\%] & 8.2 & 7.1 & 5.0 & 3.5 \\
    $T_{g,w}/T_w$ & 1.14 & 1.07 & 1.03 & 1.70 \\
    \bottomrule
  \end{tabular}
\end{table}

At the nozzle station C1 matches the edge Reynolds number of Case~3 to within 1.4\% ($\sqrt{\Rey_{x,e}}=79.2$ against 78.1), with the same wall condition (isothermal, full accommodation) and a nearly identical gas. C3 matches its edge pressure ratio (3.38 against 3.39), and C2 and C3 extend the comparison to higher Reynolds number ($\sqrt{\Rey_{x,e}}=112$ and 179). The flows differ in several respects, though. The present Mach number is higher (11.3 to 11.8 against 9.4), and the larger edge pressure ratio of C1 and C2 (5.6 and 5.0) indicates a stronger interaction. The slip is a larger fraction of the edge velocity (5.0 to 8.2\% against 3.5\%). The wall is much less cold relative to the stagnation temperature ($T_w/T_0=0.37$ to 0.49 against 0.054), so the temperature jump is smaller in relative terms. And the leading edge is sharp. The two sets of results therefore cannot be mapped onto each other. The comparison shows only that the procedure used here, which extracts the edge state, slip and temperature jump from the same field and imposes them on a similarity solution, had already been applied to a neighbouring isothermal flow at the same edge Reynolds number. The present cases extend it to stronger interaction and higher Mach number. At the nozzle station the wall-based Knudsen number $\lambda_w/\delta_{99}$ is 0.074, 0.061 and 0.041 for C1 to C3, inside the slip range $0.01<\Kn<0.1$ adopted by Klothakis et al.

\subsection{Jet-on flow field}
\label{sec:2d_flowfield}

\begin{figure}[htbp]
  \centering
  \includegraphics[width=\linewidth]{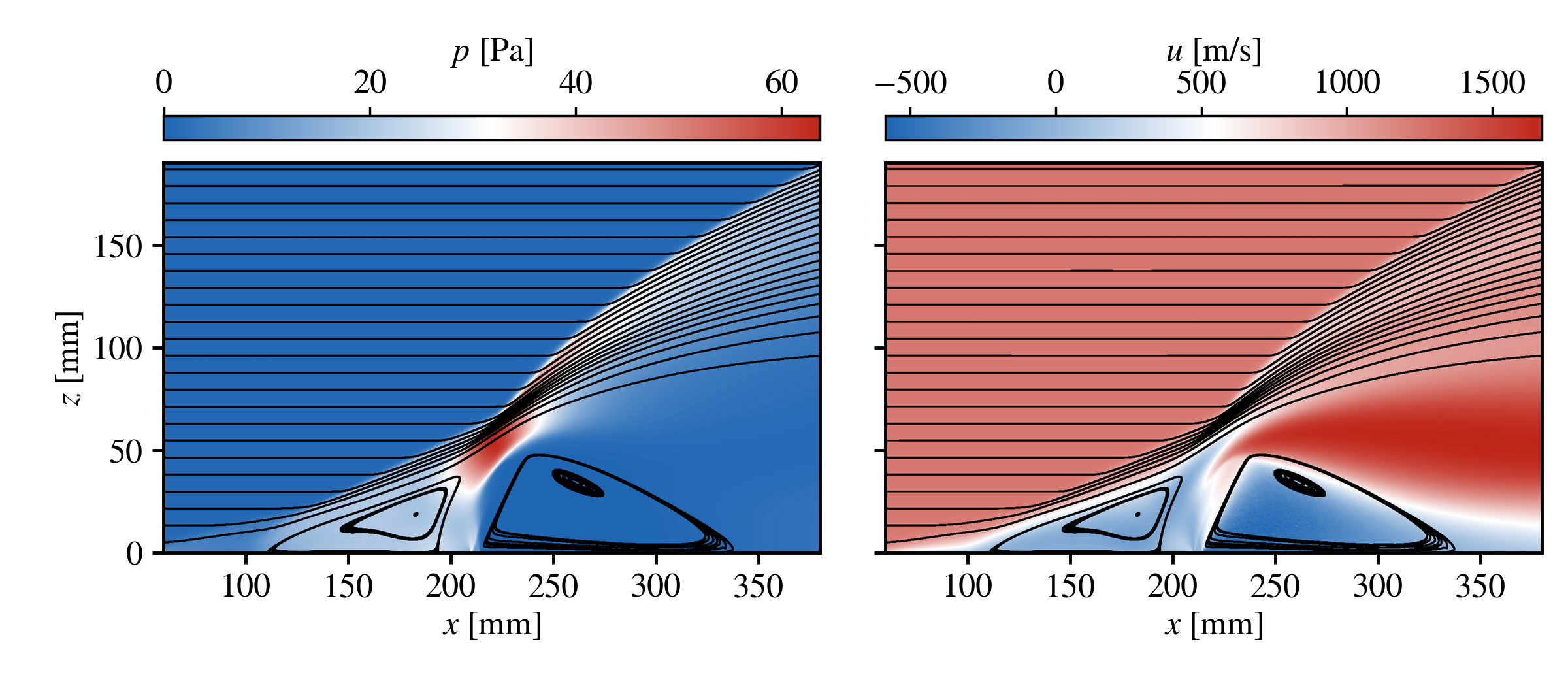}
  \caption{Planar mean-flow streamlines over the pressure (left) and the streamwise velocity (right), case C1, jet-on, in the interaction region $x=60$ to $\SI{380}{mm}$ of the $\SI{442.9}{mm}\times\SI{250}{mm}$ domain. The downstream closed region carries strongly reversed flow, and the upstream one is weak.}
  \label{fig:2d_streamlines}
\end{figure}

Figure~\ref{fig:2d_streamlines} shows mean-flow streamlines over the pressure and the streamwise velocity for C1, jet-on. The leading-edge shock is a weak compression that starts near $x=\SI{50}{mm}$. The much stronger jet bow shock and barrel shock converge on the nozzle near $x=\SI{210}{mm}$, and above the interaction the shocks merge into a single oblique front, as in Figure~\ref{fig:RCSfig}.

The streamlines show two closed recirculation regions, a smaller one just upstream of the jet, drawn as two cells that meet at a saddle point, and a larger one downstream. In the right panel of Figure~\ref{fig:2d_streamlines}, the downstream region carries strongly reversed flow, with $u$ between about $-200$ and $-500$\,m/s (down to roughly $0.4\,u_\infty$ in magnitude) over most of its interior, which makes it a genuine separated bubble. The upstream region is weak, with $u$ of mixed sign and magnitude below about 100\,m/s, and its closed streamlines mostly reflect a stagnation point in the mean field. Whether either region is sustained by collisions at its own length scale (Section~\ref{sec:jicf}) would need the local breakdown parameter evaluated for the planar field, which we have not done.

Section~\ref{sec:3dref} gives the interference pressure of the planar run and compares this flow field with its three-dimensional counterpart.

\section{Three-dimensional simulations and comparison with the planar results}
\label{sec:3dref}

This section presents the three-dimensional \DSMC{} campaign at the conditions of Section~\ref{sec:method}, in the order of Section~\ref{sec:2d}. We start with the plate-alone (jet-off) wall pressure and boundary layer, then turn to the jet-on flow field, the continuum breakdown and the interference pressure, and compare each result with its planar counterpart as it appears. The companion three-dimensional study \citep{theofilis2026companion} reports the full method, the validation against every quantity Hefer measured, and the residual analysis. We digitised Hefer's diagrams with an in-house tool, calibrated figure by figure from the printed axes.

\subsection{Plate-alone wall pressure}
\label{sec:validation}

\begin{figure}[htbp]
  \centering
  \includegraphics[width=\linewidth]{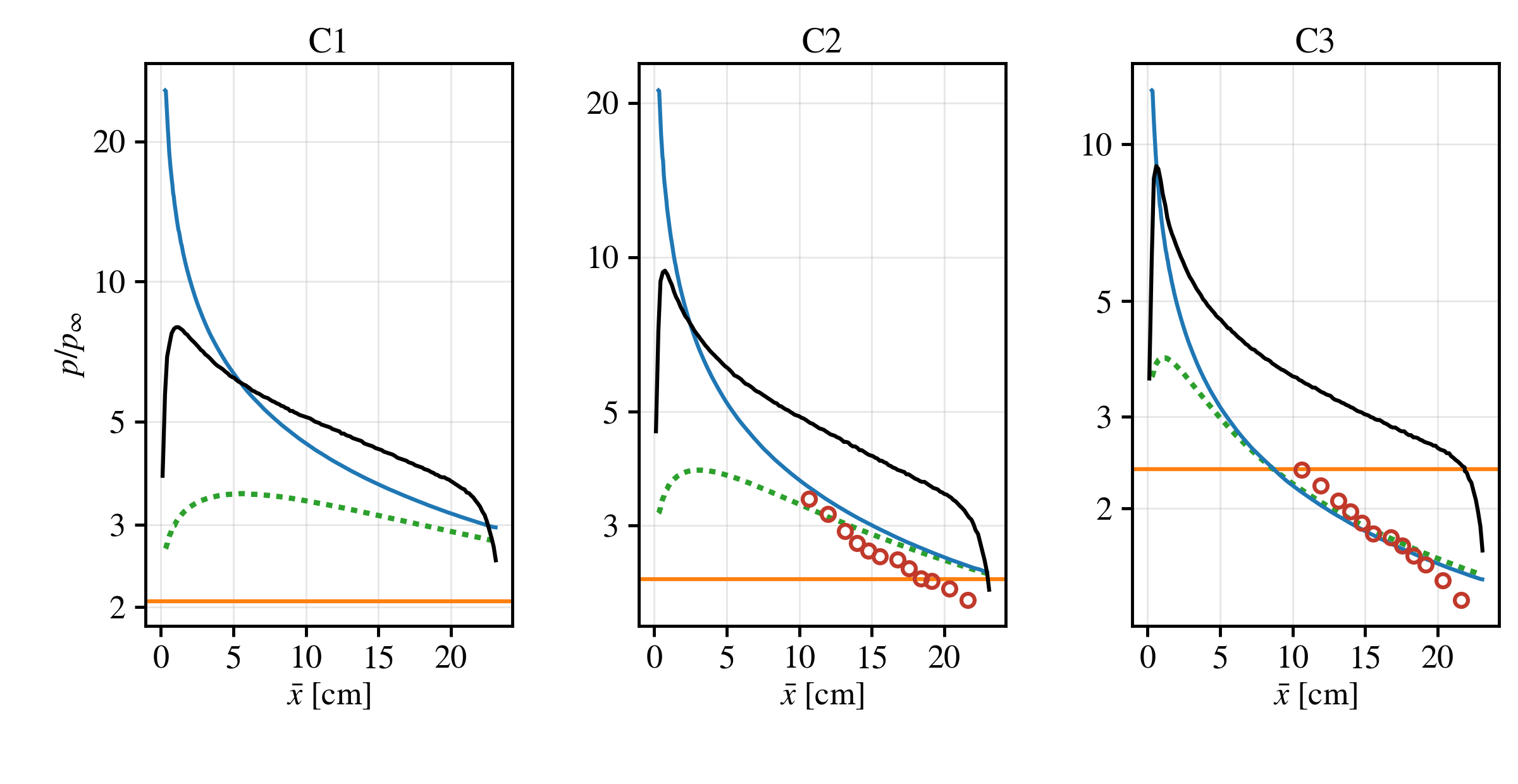}
  \caption{Plate-alone wall pressure of the three-dimensional runs (asymmetric domain) for C1, C2 and C3 (left to right). The black line shows the \DSMC{} result and the red circles Hefer's measurements (C2 and C3). The blue, orange and green dotted lines are the Cheng (continuum), Schaaf--Chambre (free-molecule) and Kienappel (blend) correlations, as in Figure~\ref{fig:2d_wallpressure}. The vertical axes are logarithmic.}
  \label{fig:val_wallpressure}
\end{figure}

Figure~\ref{fig:val_wallpressure} compares the plate-alone wall pressure with the Cheng, Schaaf--Chambre and Kienappel correlations and with Hefer's measurements, which exist for C2 and C3 only. In absolute terms the agreement is poor. At his twelve stations ($\bar x=10.6$ to $\SI{21.6}{cm}$) the three-dimensional pressure is on average 1.50 times his measurement for C2 (range 1.41 to 1.54) and 1.61 times for C3 (1.45 to 1.78). Because $\cpi$ is formed against the simulated jet-off pressure (Eq.~\ref{eq:cpi}), this offset does not enter the interference comparison directly.

\label{sec:pp_wallpressure} At Hefer's stations the planar and three-dimensional wall pressures agree with each other to within about 10\% on average (mean ratio 1.13 for C2 and 1.09 for C3, single stations up to 1.28). Their offsets from his measurements are 1.70 and 1.75 in two dimensions (Section~\ref{sec:2d_wallpressure}) and 1.50 and 1.61 in three. Both \DSMC{} models therefore share the offset from experiment. It comes from the model or its set-up, not from the dimensionality, and its cause is not established.

The two \DSMC{} models differ near the leading edge. Beyond $\bar x\approx\SI{5}{cm}$ they agree to within about 20\% at C1 and C2 and 10\% at C3, with the planar pressure the higher (the digitisation of the logarithmic axis is uncertain by about 10\%). Closer in, the planar pressure is higher still, by factors of 1.2 to 2.0 at $\bar x=\SI{1}{cm}$ and 2.4 to 4.6 at $\SI{2}{mm}$. The cell width does not explain this. The three-dimensional wall cells are $\SI{0.25}{mm}$ wide, at most 0.4 mean free paths at C1 and 0.9 at C2, and up to 1.7 at C3 over the first $\SI{6}{mm}$. The wall-adjacent gas pressure of the three-dimensional runs reproduces the planar curve to within 6\% from $\bar x=\SI{3}{mm}$ to $\SI{51}{mm}$ at C1 and C2. It lies 2.6 to 3.2 times above the three-dimensional wall-pressure curve at $\bar x=\SI{1}{mm}$, which suggests that the two curves show different quantities near the leading edge. These plate-alone differences are small beside the order-of-magnitude discrepancy in the jet-induced pressure.

\subsection{Plate-alone boundary layer}
\label{sec:2d_boundarylayer}

\begin{figure}[htbp]
  \centering
  \includegraphics[width=\linewidth]{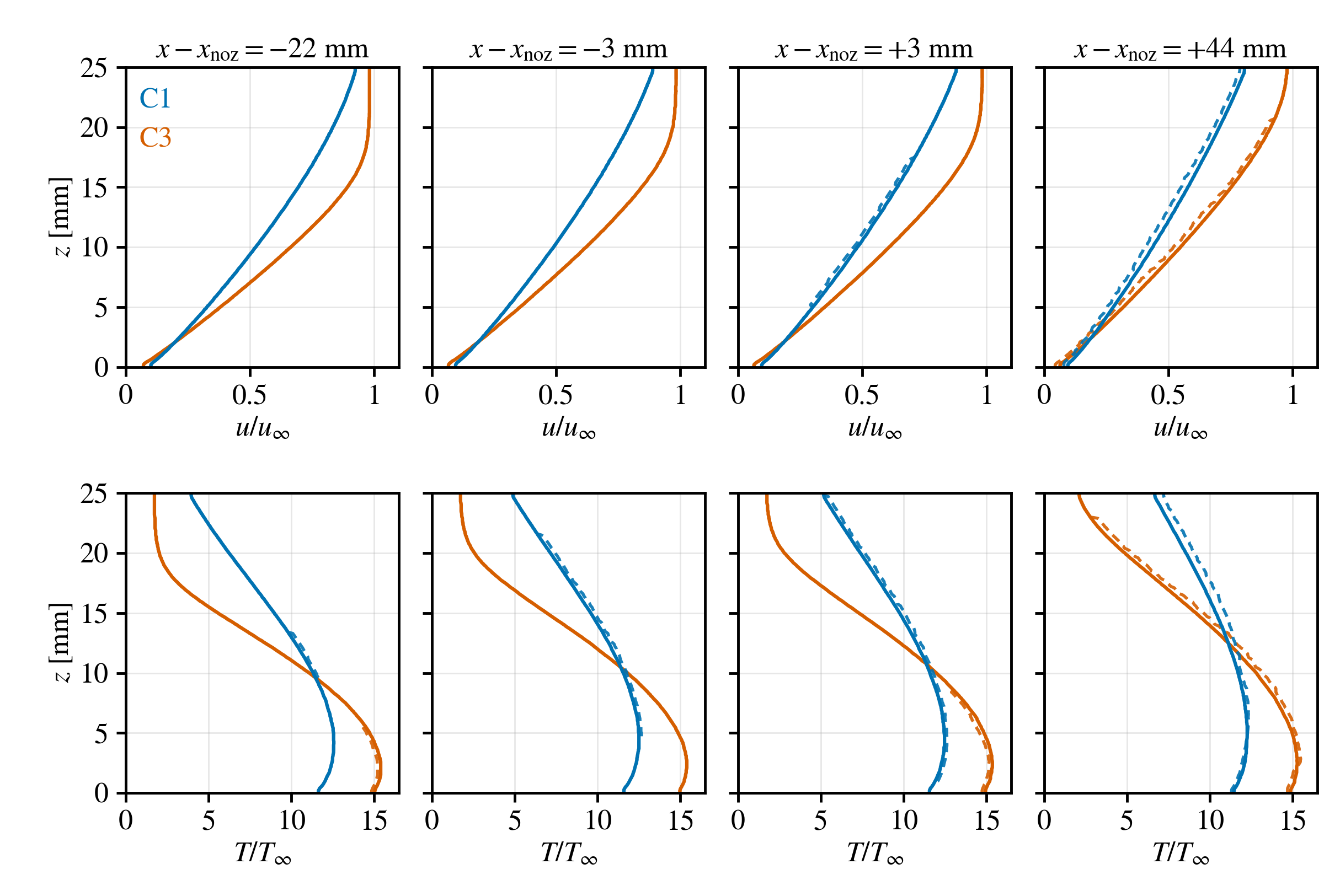}
  \caption{Plate-alone profiles of streamwise velocity (upper row) and temperature (lower row) at four stations relative to the nozzle, jet-off flow, for C1 (blue) and C3 (orange). Dashed lines show the planar run and solid lines the three-dimensional run.}
  \label{fig:bl3d}
\end{figure}

The planar and three-dimensional jet-off runs allow a direct comparison of the plate-alone boundary layer. We extracted velocity and temperature profiles from both at thirteen stations from $-120$ to $+\SI{44}{mm}$ from the nozzle ($\bar x=42$ to $\SI{206}{mm}$), using the same procedure for both. The cells within a $\SI{6}{mm}$ window in $x$ (and within $\SI{4}{mm}$ across the span in three dimensions) are averaged onto a uniform wall-normal grid. The profiles almost coincide at every station (Figure~\ref{fig:bl3d}). The velocity thickness $\delta_{99}$, the height where the smoothed velocity first reaches 99\% of its plateau value, is 0.8 to 3.9\% larger in three dimensions wherever the $\SI{30}{mm}$ window used for the plateau value does not limit it (Table~\ref{tab:bl3d}). Over all thirteen stations the near-wall temperature maximum agrees to within 1.1\% and the plateau velocity to within 1.2\%. The velocity in the wall bin agrees to within 5\% upstream of the nozzle and is up to 19\% higher in three dimensions at $+\SI{44}{mm}$, which is 0.01 to 0.02 of the plateau velocity.

At most of the eight stations nearest the nozzle for C1 and C2, the velocity layer merges with the shock layer inside the window and $\delta_{99}$ is limited by the window. The ratio of the two values is still a like-for-like comparison. We do not use the thermal-layer thickness, because its 1\% criterion is sensitive to the sharp temperature shoulder at C3 (differences up to 27\%). Repeating the five upstream stations with the three-dimensional jet-on field changes $\delta_{99}$ by at most 1.1\% and the temperature maximum by at most 0.4\%, so the jet has no detectable influence on the boundary layer there.

\begin{table}[htbp]
  \caption{Ratio of the three-dimensional to the planar velocity thickness $\delta_{99}$ and near-wall temperature maximum $T_\text{max}$ in the jet-off flow. The coordinate $\bar x$ is the distance from the leading edge. A dagger marks $\delta_{99}$ values within $\SI{5}{mm}$ of the $\SI{30}{mm}$ window used for the plateau velocity. Only the five stations upstream of the nozzle are listed.}
  \label{tab:bl3d}
  \centering
  \small
  \begin{tabular}{@{}rrcccccc@{}}
    \toprule
    & & \multicolumn{2}{c}{C1} & \multicolumn{2}{c}{C2} & \multicolumn{2}{c}{C3} \\
    \cmidrule(lr){3-4}\cmidrule(lr){5-6}\cmidrule(lr){7-8}
    $x-x_\text{noz}$ [mm] & $\bar x$ [mm] & $\delta_{99}$ & $T_\text{max}$ & $\delta_{99}$ & $T_\text{max}$ & $\delta_{99}$ & $T_\text{max}$ \\
    \midrule
    $-120$ & 42 & 1.039 & 0.995 & 1.027 & 1.002 & 1.026 & 1.002 \\
    $-100$ & 62 & 1.019 & 0.997 & 1.020 & 0.998 & 1.021 & 1.001 \\
    $-80$ & 82 & 1.016 & 0.998 & 1.016 & 0.999 & 1.017 & 1.001 \\
    $-60$ & 102 & 1.005$^\dagger$ & 0.992 & 1.009$^\dagger$ & 0.999 & 1.016 & 1.005 \\
    $-40$ & 122 & 1.011$^\dagger$ & 0.998 & 1.008 & 1.000 & 1.011$^\dagger$ & 1.007 \\
    \bottomrule
  \end{tabular}
\end{table}

\subsection{Jet-on flow field}
\label{sec:flowfield}

\begin{figure}[htbp]
  \centering
  \begin{subfigure}[b]{\linewidth}
    \includegraphics[width=\linewidth]{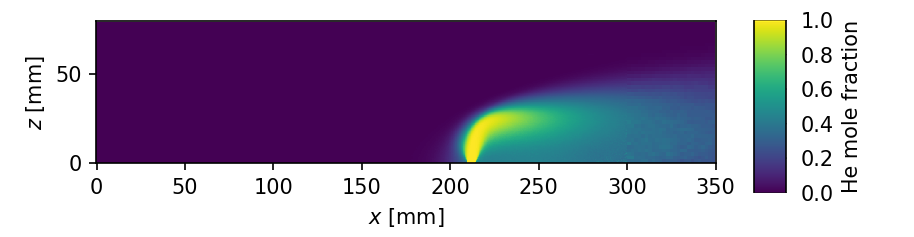}
    \caption{Helium mole fraction}
  \end{subfigure}\\[0.4em]
  \begin{subfigure}[b]{\linewidth}
    \includegraphics[width=\linewidth]{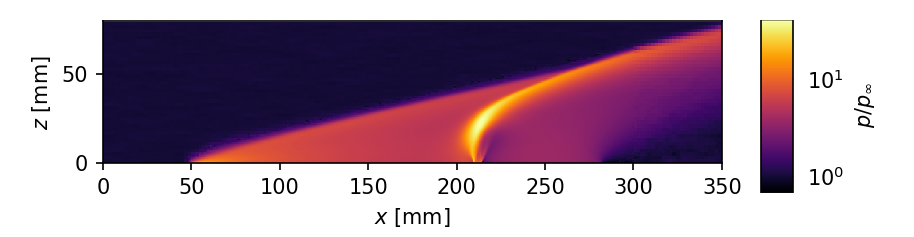}
    \caption{Pressure ratio $p/p_\infty$}
  \end{subfigure}\\[0.4em]
  \begin{subfigure}[b]{\linewidth}
    \includegraphics[width=\linewidth]{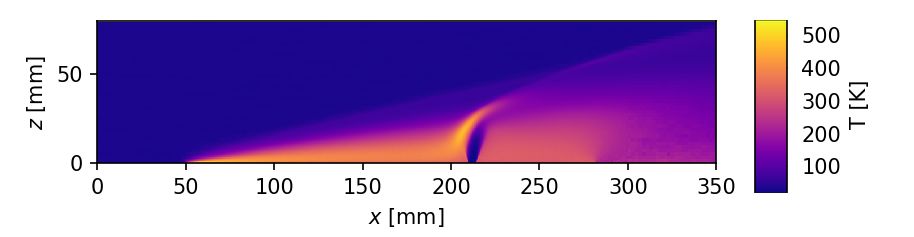}
    \caption{Temperature}
  \end{subfigure}
  \caption{Streamwise plane through the nozzle ($y=0$), case C1, jet-on.}
  \label{fig:field_sym_jeton}
\end{figure}

\begin{figure}[htbp]
  \centering
  \begin{subfigure}[b]{\linewidth}
    \includegraphics[width=\linewidth]{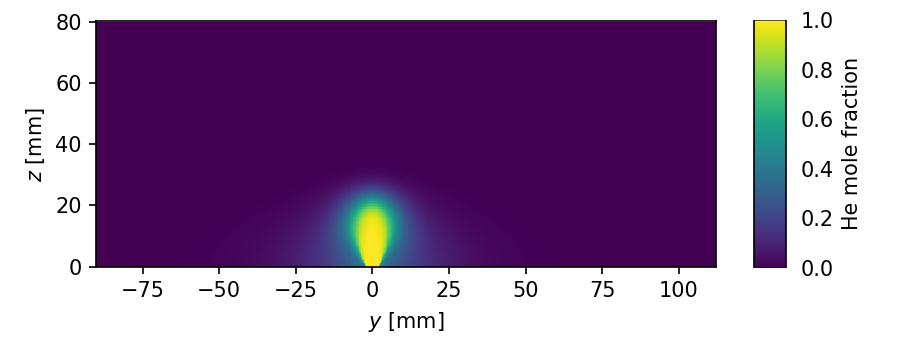}
    \caption{Helium mole fraction}
  \end{subfigure}\\[0.4em]
  \begin{subfigure}[b]{\linewidth}
    \includegraphics[width=\linewidth]{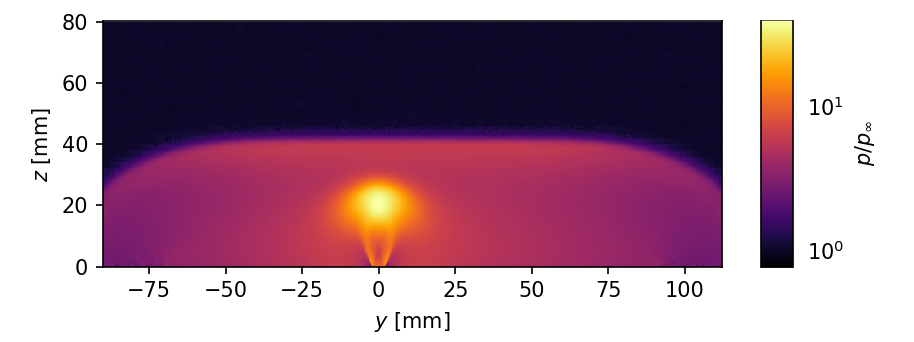}
    \caption{Pressure ratio $p/p_\infty$}
  \end{subfigure}\\[0.4em]
  \begin{subfigure}[b]{\linewidth}
    \includegraphics[width=\linewidth]{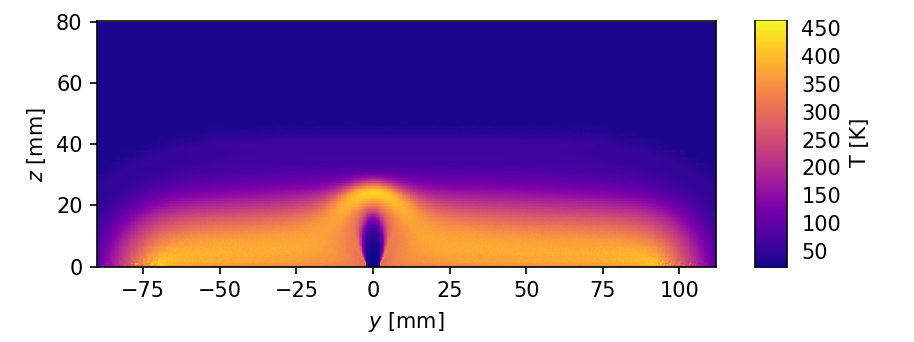}
    \caption{Temperature}
  \end{subfigure}
  \caption{Chordwise plane through the nozzle, case C1, jet-on.}
  \label{fig:field_chord_jeton}
\end{figure}

Figure~\ref{fig:field_sym_jeton} shows the streamwise plane through the nozzle ($y=0$) for C1, jet-on. A bow shock and a barrel shock converge on the nozzle, and the plume is confined to a narrow spanwise extent, with the surrounding boundary layer essentially undisturbed. Figure~\ref{fig:field_chord_jeton} shows the same case on a chordwise plane through the nozzle. The planar and three-dimensional flow fields share the same leading-edge, bow and barrel shock topology (Figure~\ref{fig:2d_streamlines}), but the disturbed region is wider and stronger in two dimensions, where the slot gives the crossflow no lateral exit. The limited spanwise extent of the plume in the chordwise view is the three-dimensional counterpart of the lateral-relief argument in Section~\ref{sec:2d_mechanism}.

\subsection{Continuum breakdown}
\label{sec:breakdown_map}

\begin{figure}[htbp]
  \centering
  \includegraphics[width=\linewidth]{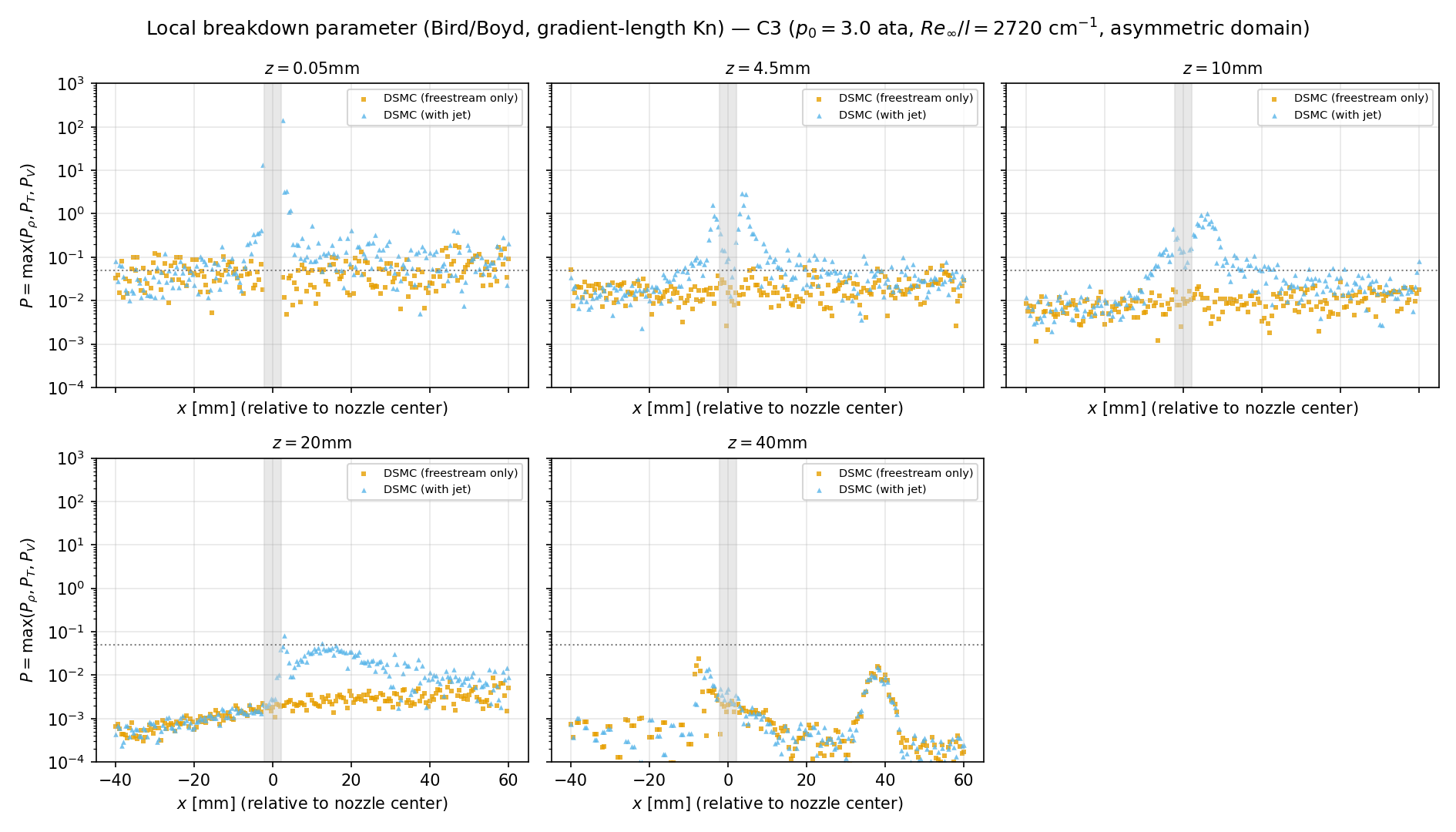}
  \caption{Local Bird--Boyd gradient-length breakdown parameter $P$, case C3, freestream-only and with jet, at five heights above the plate. The dotted line marks the commonly used breakdown threshold $P=0.05$, and the shaded band marks the nozzle's streamwise extent.}
  \label{fig:birdbreakdown}
\end{figure}

The Bird--Boyd gradient-length local Knudsen number $\Kn_\text{GLL}$
\label{sec:breakdown}
flags the regions of the interaction where the continuum assumption fails, whatever the overall rarefaction level. Figure~\ref{fig:birdbreakdown} shows the breakdown parameter $P=\max(P_\rho,P_T,P_V)$ at five heights above the plate for C3, Hefer's most continuum condition and therefore the most conservative test. Even there, the jet-on curve exceeds the commonly used threshold $P\gtrsim0.05$ by one to two orders of magnitude in the near-wall region around the nozzle, while the jet-off curve stays below it almost everywhere. The jet-induced field, not the bare plate boundary layer, drives the breakdown, so \DSMC{} is needed for the shock and shear-layer regions even at Hefer's mildest condition.

\subsection{Interference pressure of the three-dimensional reference}
\label{sec:val_cpi}

\begin{figure}[htbp]
  \centering
  \includegraphics[width=\linewidth]{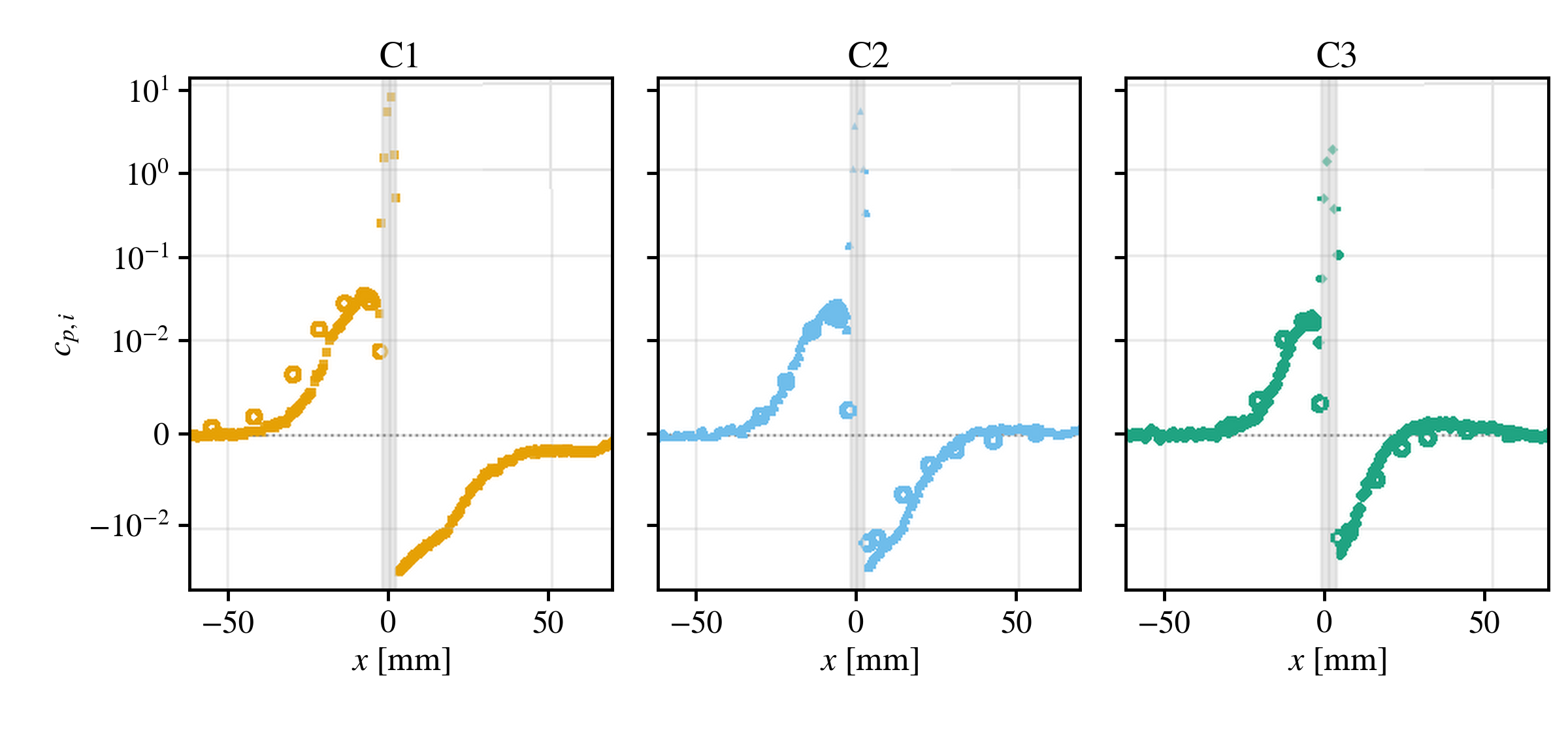}
  \caption{Interference pressure coefficient $\cpi(x)$ from three-dimensional \DSMC{} (small filled markers) and Hefer's measurements (open circles), for C1, C2 and C3 (left to right), each in its own colour. The vertical axis is symmetric-logarithmic, $x$ is measured from the nozzle centre, and the shaded band marks the nozzle exit region.}
  \label{fig:val_cpi}
\end{figure}

With the jet active, Figure~\ref{fig:val_cpi} compares $\cpi(x)$ with Hefer's measurements along the nozzle centreline. Both the blockage lobe upstream of the nozzle and the entrainment well downstream are reproduced. The lobe peak is 0.032, 0.022 and 0.017 for C1 to C3, against 0.034, 0.029 and 0.016 measured by Hefer (within 25\%), at $x=-9$, $-9$ and $\SI{-5}{mm}$ against $-10$, $-10$ and $\SI{-8}{mm}$. For C1 the sampling uncertainty of the lobe peak, from the two production windows, is below 1\% (Appendix~\ref{app:verification}). Agreement of $\cpi$ with Hefer's measurements is evidence for the jet-on response, but not an independent validation of it. Forming $\cpi$ against the simulated jet-off pressure removes the baseline offset only if the jet-induced pressure adds to the baseline, and we did not test how the interaction responds to a base state that is 50 to 60\% too high. The lobe peak agrees to within 25\%, the upstream flank of the C1 lobe is underpredicted, and we cannot separate these differences from the offset in the base state. Plausible sources of the offset are the non-parallel flow of Hefer's tunnel nozzle, which he himself invokes, the freestream state inferred from the tunnel calibration, and the full accommodation assumed at the wall, which for helium on a cooled metal surface is uncertain and which we did not test. At C2 and C3, where Hefer has downstream points, the well depth is $-0.017$ and $-0.015$ against $-0.018$ and $-0.012$. The one systematic difference is on the upstream flank of the C1 lobe, where the \DSMC{} value is 0.3 to 0.4 of Hefer's between 30 and 45 mm ahead of the nozzle and 0.6 to 0.75 between 12 and 20 mm. The companion study reports two further residuals, a small offset in the flow-angle minima and a 41.5\% deficit in off-axis helium concentration at one station. Neither affects the pressure comparisons used here.

\subsection{Planar versus three-dimensional interference pressure}
\label{sec:2d_interference}

\begin{figure}[htbp]
  \centering
  \includegraphics[width=\linewidth]{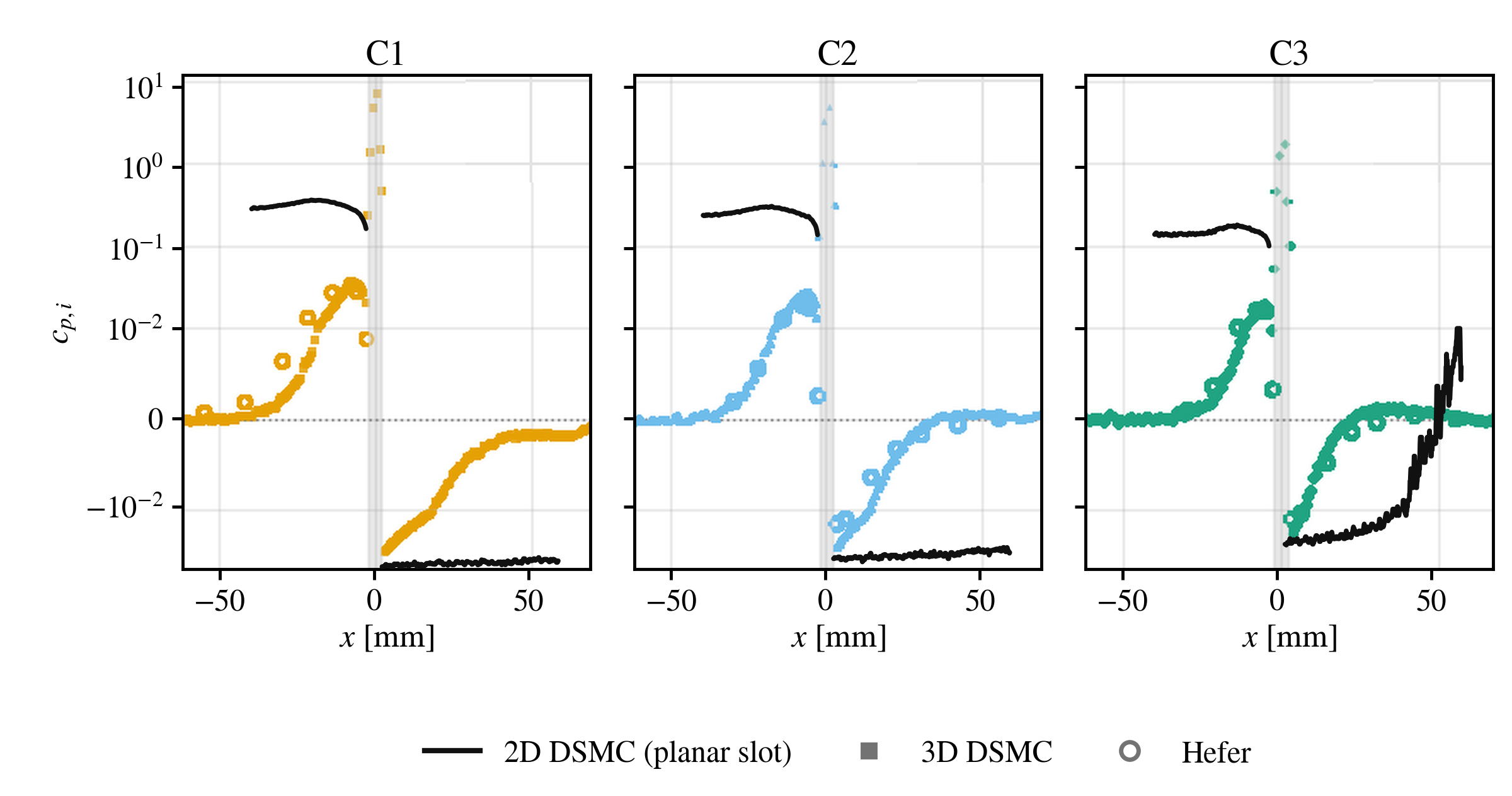}
  \caption{Interference pressure coefficient $\cpi(x)$ for C1, C2 and C3 (left to right) from the planar-slot \DSMC{} (line), the three-dimensional \DSMC{} (filled markers) and Hefer's measurements (open circles), on a common symmetric-logarithmic axis. The shaded band marks the nozzle exit region. The planar curve is omitted for $|x|<\SI{2.5}{mm}$ and plotted over $-40\le x\le\SI{60}{mm}$ only.}
  \label{fig:2d_cpi}
\end{figure}

Figure~\ref{fig:2d_cpi} compares the interference pressure coefficient from the planar slot, the three-dimensional reference and Hefer's measurements on a common symmetric-logarithmic axis, and Table~\ref{tab:cpi_metrics} gives the numbers.

The planar field shares three features with the three-dimensional one. The sign and two-lobed structure are the same, with a positive blockage lobe upstream of the nozzle and a negative entrainment well downstream. The lobe peak falls from C1 to C3 in both, from 0.394 to 0.192 in two dimensions and from 0.032 to 0.017 in three. And the well recovers sooner in the denser case, with the planar C3 well returning to zero at about $\SI{52}{mm}$.

It differs in three ways. The first is the magnitude. The planar lobe peak is 0.394, 0.329 and 0.192 for C1 to C3. That is 11 to 15 times the three-dimensional value (0.032, 0.022 and 0.017) and 11.7, 11.5 and 11.8 times Hefer's measured peak (0.034, 0.029 and 0.016). The ratio to experiment is constant to within 3\% across the three conditions, whereas the three-dimensional peaks are within 25\% of his. The second is the upstream extent. At $x=\SI{-40}{mm}$, about nine nozzle diameters ahead of the nozzle, the planar coefficient is still 0.315, 0.254 and 0.151, close to 80\% of its peak. The three-dimensional value there is below $10^{-3}$, and Hefer's is $2\times10^{-3}$ or less. The planar lobe also peaks farther from the nozzle, at $-20$, $-18$ and $\SI{-13}{mm}$, against $-9$, $-9$ and $\SI{-5}{mm}$ in three dimensions and about $\SI{-9}{mm}$ in Hefer's data. The third is the entrainment well. It is two to three times deeper than in three dimensions ($-0.057$, $-0.047$ and $-0.029$ against $-0.030$, $-0.017$ and $-0.015$), and for C1 and C2 it is still at about 80\% of its depth $\SI{50}{mm}$ downstream, where the three-dimensional well and Hefer's data have recovered to within $2\times10^{-3}$.

The difference in jet density (Section~\ref{sec:2d_config}) explains only a small part of the factor. The planar runs satisfy the same resolution and sampling checks as the three-dimensional ones, and the overprediction does not change with resolution across the three conditions (Appendix~\ref{app:verification}), so it is not a resolution artefact (Section~\ref{sec:2d_mechanism}).

\begin{table}[htbp]
  \centering
  \caption{Metrics of the interference pressure coefficient for planar \DSMC{} (2D), three-dimensional \DSMC{} (3D) and Hefer's measurements (H). The coordinate $x$ is measured from the nozzle centre in millimetres, and the well minimum is the most negative smoothed value downstream of the nozzle. The 2D and 3D values are read from the plotted curves. Hefer's values are from his digitised data, and he has no downstream points at C1.}
  \label{tab:cpi_metrics}
  \footnotesize
  \setlength{\tabcolsep}{2.4pt}
  \begin{tabular}{@{}lccccccccc@{}}
    \toprule
    & \multicolumn{3}{c}{C1} & \multicolumn{3}{c}{C2} & \multicolumn{3}{c}{C3} \\
    \cmidrule(lr){2-4}\cmidrule(lr){5-7}\cmidrule(lr){8-10}
    & 2D & 3D & H & 2D & 3D & H & 2D & 3D & H \\
    \midrule
    Lobe peak $\cpi$ & 0.394 & 0.032 & 0.034 & 0.329 & 0.022 & 0.029 & 0.192 & 0.017 & 0.016 \\
    Ratio 2D/H       & \multicolumn{3}{c}{11.7} & \multicolumn{3}{c}{11.5} & \multicolumn{3}{c}{11.8} \\
    Peak position    & $-20$ & $-9$ & $-10$ & $-18$ & $-9$ & $-10$ & $-13$ & $-5$ & $-8$ \\
    $\cpi(x=-40)$    & 0.315 & $<10^{-3}$ & 0.002 & 0.254 & $<10^{-3}$ & $<10^{-3}$ & 0.151 & $<10^{-3}$ & $<10^{-3}$ \\
    Well minimum     & $-0.057$ & $-0.030$ & n.a. & $-0.047$ & $-0.017$ & $-0.018$ & $-0.029$ & $-0.015$ & $-0.012$ \\
    Well position    & 6 & 3 & n.a. & 12 & 6 & 3 & 3 & 4 & 3 \\
    $\cpi(x=+50)$    & $-0.046$ & $-0.002$ & n.a. & $-0.036$ & 0.001 & $-0.001$ & $-0.003$ & 0.001 & 0.001 \\
    \bottomrule
  \end{tabular}
\end{table}

\subsection{Summary of the comparison}
\label{sec:2d_summary}

Table~\ref{tab:survive} collects the comparison. The planar slot reproduces the qualitative structure of the interaction and its trend with rarefaction, but it gets the magnitude and extent of the surface pressure wrong.

\begin{table}[htbp]
  \centering
  \caption{Comparison of the planar slot with the three-dimensional reference, metric by metric.}
  \label{tab:survive}
  \small
  \setlength{\tabcolsep}{4pt}
  \begin{tabular}{@{}>{\raggedright\arraybackslash}p{3.4cm}>{\raggedright\arraybackslash}p{5.3cm}>{\raggedright\arraybackslash}p{2.9cm}@{}}
    \toprule
    Metric & Two-dimensional slot & Verdict \\
    \midrule
    Shock topology & Leading-edge, jet bow and barrel shocks present; disturbed region wider and stronger & Qualitatively reproduced \\
    Sign and structure of $\cpi$ & Upstream blockage lobe and downstream entrainment well in both & Reproduced \\
    Ordering with rarefaction & Lobe peak falls from C1 to C3 in both & Reproduced \\
    Peak of blockage lobe & 0.39, 0.33 and 0.19: 11 to 15 times the three-dimensional values and 11.5 times Hefer's, at every condition & Not reproduced \\
    Upstream extent of lobe & About 80\% of peak at $\SI{-40}{mm}$, against $<10^{-3}$ in three dimensions & Not reproduced \\
    Entrainment well & Two to three times deeper; no recovery within $\SI{60}{mm}$ for C1 and C2 & Not reproduced \\
    Plate-alone wall pressure & Within about 20\% of the three-dimensional value beyond $\bar x\approx\SI{5}{cm}$ (three-dimensional about 50 to 60\% and planar about 70 to 75\% above Hefer's data); planar higher within 2 cm of the leading edge & Approximately reproduced \\
    \bottomrule
  \end{tabular}
\end{table}


\section{Origin of the discrepancy}
\label{sec:2d_mechanism}

Section~\ref{sec:2d_interference} shows that the planar slot reproduces the structure of the interference pressure and its ordering with rarefaction, but overpredicts the peak by roughly an order of magnitude and carries it much farther upstream. Here we ask which physical mechanism is responsible, and argue that the gap comes from the geometry, not the numerics.

One might expect blockage, which relies on an obstacle producing a bow shock in the oncoming flow, to be less sensitive to dimensionality, and entrainment, a sink-like inflow that Hefer models as axisymmetric, to carry the effect. In fact the largest discrepancy, a factor of about ten, is in the blockage lobe, while the entrainment well is deeper and longer-lived in two dimensions by a smaller factor. The blockage lobe is the same feature in both geometries, with the same sign, the same position relative to the nozzle and the same trend with rarefaction. Only its size and extent differ.

The explanation we propose is lateral relief. A planar slot is an unbroken spanwise obstacle, so the crossflow can only go over it. A round jet lets the crossflow divert around it, and that diversion relieves the upstream compression. The spanwise velocity field shows the diversion directly. The picture also agrees with Powrie's slot-jet experiments, which found that the size of the obstacle the jet presents to the oncoming flow sets the main characteristics of the upstream interaction \citep{powrie1996}.

\begin{figure}[htbp]
  \centering
  \includegraphics[width=\linewidth]{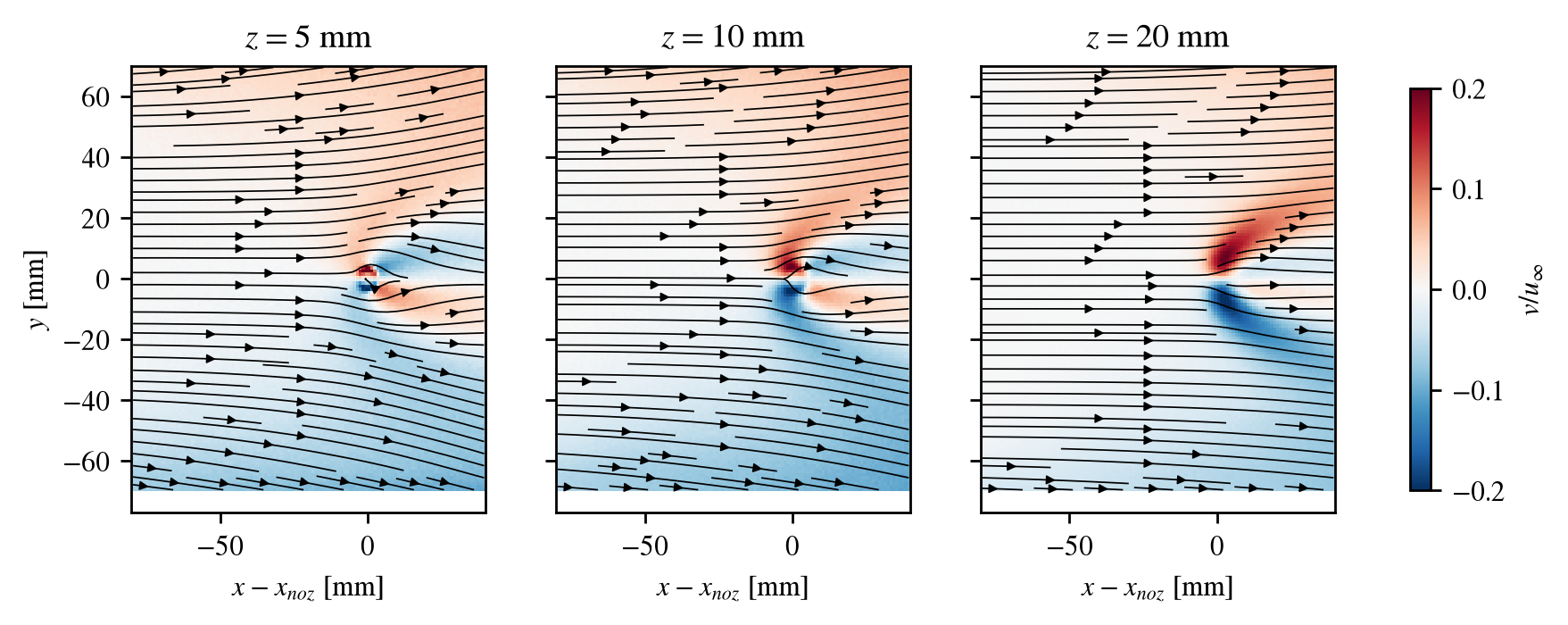}\\[3pt]
  \includegraphics[width=\linewidth]{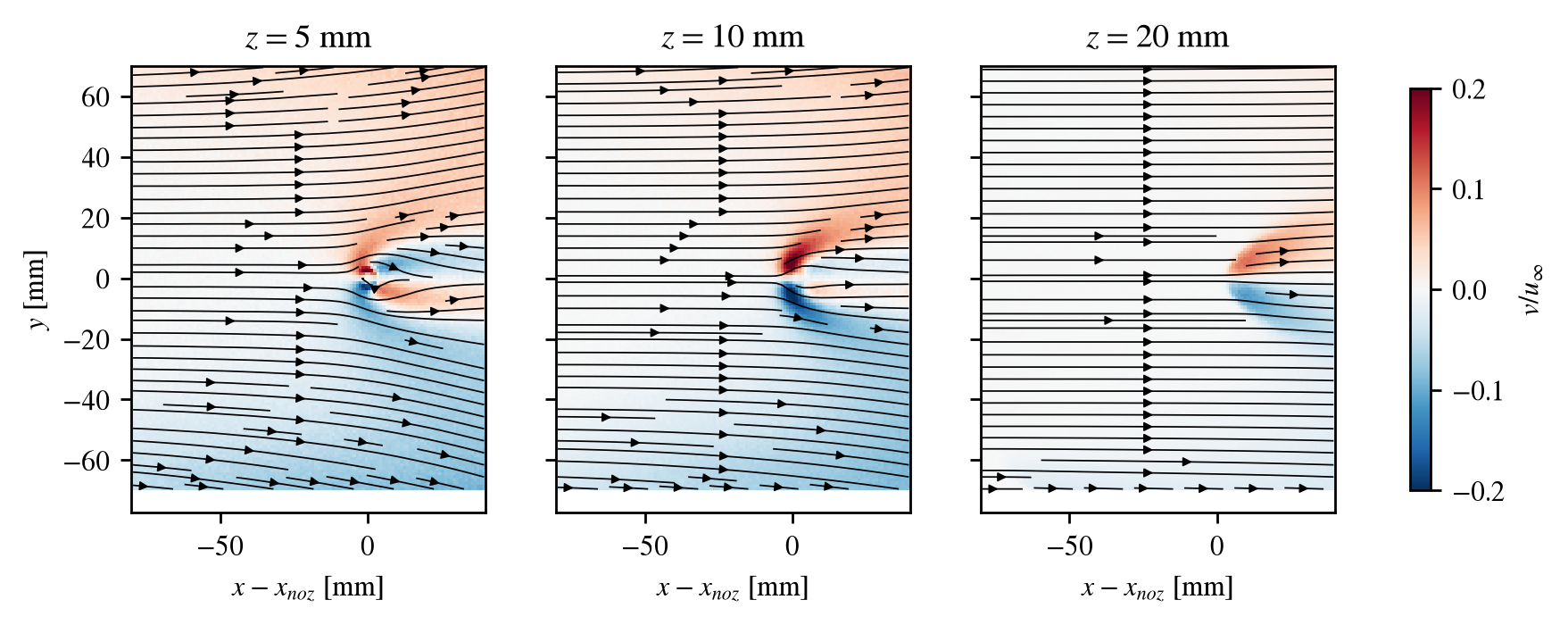}\\[3pt]
  \includegraphics[width=0.85\linewidth]{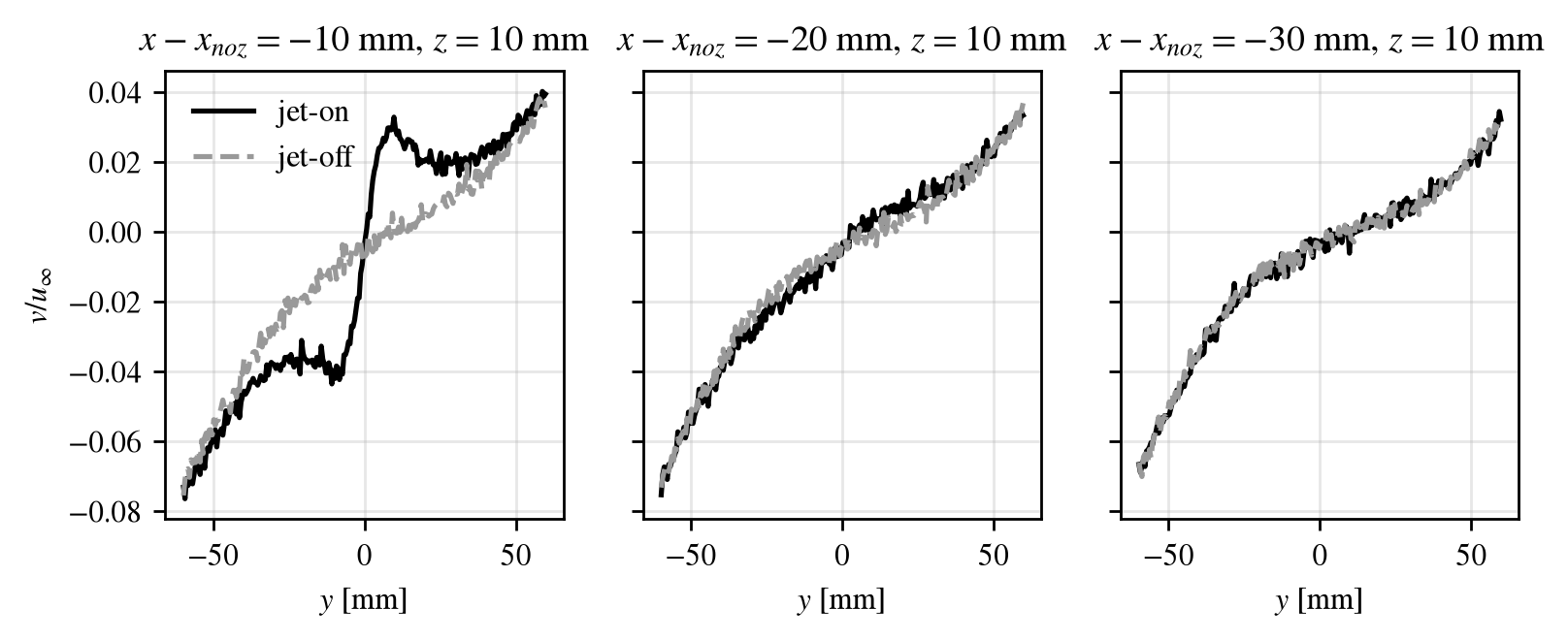}
  \caption{Spanwise velocity $v/u_\infty$ (colour) with streamlines in planes parallel to the wall at $z=5$, 10 and $\SI{20}{mm}$, from the three-dimensional jet-on runs for C1 (top) and C3 (middle). The nozzle is at the origin and the flow is from left to right. The bottom row shows the spanwise velocity at $z=\SI{10}{mm}$ for C1 at three stations ahead of the nozzle, with the jet on (solid) and off (dashed).}
  \label{fig:relief}
\end{figure}

Figure~\ref{fig:relief} shows the spanwise velocity $v$ in planes parallel to the wall for C1 and C3, from the three-dimensional runs with the jet on. Ahead of the nozzle the near-wall flow turns away from the axis, and the streamlines open out from about $\SI{15}{mm}$ upstream. The pattern is weaker at C3 and at greater height. The profiles in the lower panel compare the spanwise velocity at $z=\SI{10}{mm}$ with and without the jet. Beyond $\SI{20}{mm}$ ahead of the nozzle they coincide, and the smooth drift they share is the lateral spreading of the finite plate. At $x-x_\text{noz}=\SI{-10}{mm}$ the jet-on velocity changes by about $0.06$ to $0.07\,u_\infty$ across the axis at C1 and C2 and by $0.02\,u_\infty$ at C3. The jet-off field changes by only about $0.015\,u_\infty$ at C1 and C2 and $0.01\,u_\infty$ at C3 over the same span.

To measure how much flow the jet diverts, Table~\ref{tab:relief} gives the net sideways number flux through the planes $y=\pm Y'$ between $\SI{40}{mm}$ and $\SI{3}{mm}$ ahead of the nozzle, at fixed height $z$. We divide it by the flux entering over $|y|<Y'$ at $\SI{40}{mm}$ ahead, and subtract the same quantity from the jet-off field. In the layer closest to the wall ($z=\SI{5}{mm}$), the jet turns aside 33 to 40\% of the flow aimed at a strip $\SI{10}{mm}$ wide (about two nozzle diameters) before it reaches the nozzle, at all three conditions. At $z=\SI{10}{mm}$ the fraction is 44\% and 41\% at C1 and C2 and 13\% at C3, and at $\SI{20}{mm}$ it falls to 6\%, 2\% and zero. For a slot every one of these fractions is zero by construction. The diversion is concentrated in the near-wall layer, and it becomes thinner and weaker as the freestream density rises and $J$ falls from C1 to C3.

The subtraction matters because the plate spreads the flow sideways by itself. For strips up to $\SI{10}{mm}$ wide the jet-off fraction is 2.6 to 5.0\% at $z=\SI{5}{mm}$, 1.0 to 2.7\% at $\SI{10}{mm}$ and below 0.6\% at $\SI{20}{mm}$. Over the whole $\SI{40}{mm}$ column and the $\SI{30}{mm}$ half-width the plate accounts for 77 to 83\% of the jet-on value, which is why we quote the diversion for the near-wall layer and narrow strips. The measurement shows that the diversion exists and how large it is. Whether relief explains the whole factor of about 12 would take a planar calculation to which relief is added.

\begin{table}[htbp]
  \caption{Share of the incoming flow turned sideways by the jet ahead of the nozzle, in per cent. It is the net sideways number flux through the planes $y=\pm Y'$ between $\SI{40}{mm}$ and $\SI{3}{mm}$ ahead of the nozzle, divided by the flux entering over $|y|<Y'$ at $\SI{40}{mm}$ ahead, at fixed height $z$, with the same quantity from the jet-off field subtracted. It is zero for a planar slot by construction.}
  \label{tab:relief}
  \centering
  \small
  \begin{tabular}{@{}lcccccc@{}}
    \toprule
    & \multicolumn{3}{c}{$Y'=\SI{5}{mm}$} & \multicolumn{3}{c}{$Y'=\SI{10}{mm}$} \\
    \cmidrule(lr){2-4}\cmidrule(lr){5-7}
    $z$ [mm] & C1 & C2 & C3 & C1 & C2 & C3 \\
    \midrule
    5  & 38 & 40 & 33 & 19 & 21 & 13 \\
    10 & 44 & 41 & 13 & 14 & 13 & 2.8 \\
    20 & 6.1 & 2.1 & 0.0 & 1.9 & 0.5 & 0.0 \\
    \bottomrule
  \end{tabular}
\end{table}

Without a lateral direction, the real flow loses its main route for relieving the blockage. The upstream pressure rise is then larger and, with nowhere to go, extends farther ahead of the obstacle. Entrainment suffers from the same missing direction. Hefer's plane sink flow draws its mass from all azimuths around the jet, and a slot has none to draw from. This is a secondary effect on the numbers above.

We have not isolated lateral relief as the cause. Besides dimensionality, the two jets differ in exit density (the planar jet is 1.6 times denser), in exit size (a slot of $\SI{4.0}{mm}$ against a circular exit of $\SI{4.3}{mm}$), in divergence (a parallel stream against a conical source of $\SI{8.5}{\degree}$ half-angle) and in the plate (zero thickness and unlimited span against a bevelled plate of finite span). Only the density difference is estimated, as a factor of 1.6 if the interference scales linearly with jet strength. The measured diversion of the flow shows that relief exists and is large, and a slot cannot provide it. We did not run a planar case with relief added, or a three-dimensional case with the other differences removed, so relief remains our proposed explanation for the rest of the factor and not a demonstrated cause. These arguments also rest on time-averaged fields, and unsteady relief may differ (Section~\ref{sec:conclusions}).

Two observations point to a geometric effect, not a numerical one. First, the planar runs satisfy the resolution and sampling checks of Appendix~\ref{app:verification}, and the overprediction is the same, 11.5 to 11.8 times Hefer's peak, at three densities whose cell-to-mean-free-path ratios differ six-fold. The distribution has the right structure but the wrong scale. A missing degree of freedom would produce that, whereas under-resolution would not. Second, consider the base flow. The plate-alone boundary layer agrees with its three-dimensional counterpart to within 4\% in thickness, where the averaging window does not limit it, from $\SI{120}{mm}$ ahead of the nozzle to $\SI{44}{mm}$ behind it (Section~\ref{sec:2d_boundarylayer}), and the plate-alone wall pressure (Section~\ref{sec:pp_wallpressure}) agrees to within about 10\% on average. Yet the surface pressure induced by the jet is wrong by a factor of ten. The same geometric reduction costs tens of percent in one quantity and an order of magnitude in the other.

\section{Conclusions and outlook}
\label{sec:conclusions}

We assessed a planar slot jet as a \DSMC{} model of reaction-control-jet interference in a rarefied hypersonic crossflow, against Hefer's 1975 measurements and a three-dimensional \DSMC{} reference.

The three-dimensional reference reproduces the measured interference pressure. The lobe peak is within about 25\% of Hefer's at all three conditions (0.032, 0.022 and 0.017 against 0.034, 0.029 and 0.016), its position within 3\,mm, and the well depth within about 25\% where he measured it. Its plate-alone wall pressure is 50 to 60\% above his measurements at C2 and C3, and the planar runs show a similar offset (70 to 75\%) whose cause is not established. Because $\cpi$ is formed against the simulated jet-off pressure, the interference comparison is less exposed to this offset than the base flow is, although this does not by itself validate the jet-on response. In the three-dimensional field the Bird--Boyd breakdown parameter exceeds the usual threshold by one to two orders of magnitude near the wall around the nozzle with the jet on, even at the most continuum condition, so a kinetic treatment is needed there.

The planar slot reproduces the qualitative shock topology, the plate-alone boundary layer (within 4\% in thickness, where the averaging window does not limit it, and within about 1\% in peak temperature, from $\SI{120}{mm}$ ahead of the nozzle to $\SI{44}{mm}$ behind it), the plate-alone wall pressure (within about 10\% of the three-dimensional value on average at Hefer's stations, and up to 28\% at single stations), the sign and two-lobed structure of the interference pressure, and the ordering of the lobe peak with rarefaction. At the nozzle station the plate-alone boundary layer is also consistent with a similarity solution that carries the slip velocity and temperature jump of the \DSMC{} field, to 2 to 3\% rms in velocity against 4 to 8\% for the no-slip solution. The procedure is the one that Klothakis et al.\ \citep{Klothakis2022} applied to an isothermal flow at a comparable edge Reynolds number. It misses both the magnitude and the extent of the jet-induced surface pressure. The peak of the blockage lobe is $\cpi\approx0.39$ at the most rarefied condition and $\approx0.19$ at the most continuum one, 11 to 15 times the three-dimensional values. The lobe has still not decayed $\SI{40}{mm}$ (about nine nozzle diameters) ahead of the nozzle, where the three-dimensional perturbation is below $10^{-3}$. The downstream entrainment well is two to three times deeper and, at the two more rarefied conditions, does not recover within $\SI{60}{mm}$.

We propose that the discrepancy is mainly due to the missing lateral relief (Section~\ref{sec:2d_mechanism}). A slot is an unbroken spanwise obstacle that the crossflow can only pass over, while a round jet lets the crossflow divert around it. The three-dimensional spanwise velocity shows that it does. At $\SI{5}{mm}$ above the wall the jet turns aside 33 to 40\% of the flow aimed at a strip two nozzle diameters wide ahead of the nozzle, at all three conditions (Table~\ref{tab:relief}), whereas a slot turns none aside. The denser planar jet accounts for about a factor of 1.6 of the 11 to 15 times difference in the peak if the interference scales linearly with jet strength, which leaves a factor of 7 to 9 for the geometry. The discrepancy is largest in the blockage lobe and smaller in the entrainment well, although Hefer's model makes entrainment axisymmetric and might lead one to expect the opposite. Whether relief explains the whole factor would take a planar calculation to which relief is added.

A planar calculation can therefore show the direction in which a change of freestream density moves the interaction, and it is adequate for quantities that do not depend on lateral mass exchange. For the jet-induced surface pressure, and so for the interference force a designer needs, a planar result must be calibrated against three-dimensional data. A planar idealisation should become defensible only when the spanwise extent of the jet exceeds the upstream reach of the planar interference, about $\SI{40}{mm}$ or nine nozzle diameters at C1, where the planar lobe is still 80\% of its peak. This is an observation from these three cases rather than a general threshold, and we did not test it.

These findings come with limitations. They hold for one geometry and one gas pair, a flat plate with helium injected into air, and for steady time-averaged solutions. Freestream density is the only parameter we varied, at one fixed jet state, so the trend with jet strength is untested. The planar jet is 1.6 times denser than the three-dimensional jet (Section~\ref{sec:2d_config}), which muddies part of the comparison. We argue full thermal accommodation at the wall from Hefer's water-cooled copper plate but did not test it. Instantaneous drag deviations of order 100\% have been reported in a comparable configuration \citep{Karpuzcu2023}, and we have not established whether the interference pressure is as sensitive. The shock interiors are resolved only to a compromise (about four mean free paths in the finest cells of the planar C3 post-shock region), and vibrational excitation and chemistry are switched off. We did not compute a breakdown map for the planar recirculation regions or a formal self-similarity collapse of the boundary-layer profiles, and the digitised data carry about 10\% uncertainty from reading logarithmic axes.

Two extensions follow. The digitised Hefer data set is worth publishing in its own right, since the original report survives only as printed diagrams and a machine-readable form makes a fifty-year-old benchmark usable by anyone. A companion three-dimensional study \citep{theofilis2026companion} extends the validated configuration by two and a half decades of Knudsen number. Whether the planar discrepancy found here narrows or widens across that range is an open question that this paper does not address. The question matters because the interference is large. In the three-dimensional simulations of the companion study it is 36\% to 117\% of the jet's own thrust \citep{theofilis2026companion}, and nothing guarantees that its sign or its point of application helps the intended manoeuvre.

\begin{appendix}
\section{Species and collision parameters}
\label{app:vss}

Table~\ref{tab:app_vss} gives the variable soft sphere parameters and species data used in every simulation. The helium entry is Bird's variable hard sphere diameter \citep{bird1994} with $\alpha=1$. The N$_2$ and O$_2$ entries differ slightly from Bird's variable soft sphere table (N$_2$: $d=\SI{4.11}{\angstrom}$, $\alpha=1.36$; O$_2$: $d=\SI{4.01}{\angstrom}$, $\alpha=1.40$), which makes the viscosity of N$_2$ and O$_2$ in the simulations about 2.3\% and 2.5\% higher than Bird's, too little to matter for the results. Cross-species pairs take the \SPARTA{} default, the arithmetic mean of the parameters of the two species, so that the helium--nitrogen diameter is \SI{3.20}{\angstrom}. Helium has no rotational degrees of freedom, so Borgnakke--Larsen exchange applies only to the two air species, and vibrational relaxation is switched off at run time (frozen chemistry).

\begin{table}[htbp]
  \centering
  \caption{VSS collision parameters and species data. $d_\text{ref}$ is the
  reference diameter at $T_\text{ref}$, $\omega$ the viscosity temperature
  exponent, and $\alpha$ the VSS scattering exponent ($\alpha=1$ recovers
  variable hard sphere). $\zeta_\text{rot}$ is the number of rotational
  degrees of freedom and $\Theta_\text{vib}$ the characteristic vibrational
  temperature.}
  \label{tab:app_vss}
  \small
  \begin{tabular}{@{}lcccccccc@{}}
    \toprule
    & $d_\text{ref}$ & $\omega$ & $T_\text{ref}$ & $\alpha$
    & $M$ & $m$ & $\zeta_\text{rot}$ & $\Theta_\text{vib}$ \\
    & [m] & & [K] & & [amu] & [kg] & & [K] \\
    \midrule
    N$_2$ & \num{4.07e-10} & 0.74 & 273.15 & 1.6 & 28.016 & \num{4.65e-26}  & 2 & 3371 \\
    O$_2$ & \num{3.96e-10} & 0.77 & 273.15 & 1.4 & 32.00  & \num{5.31e-26}  & 2 & 2256 \\
    He    & \num{2.33e-10} & 0.66 & 273.15 & 1.0 & 4.003  & \num{6.646e-27} & 0 & n.a. \\
    \bottomrule
  \end{tabular}
\end{table}

\section{Numerical verification}
\label{app:verification}

This appendix states what was checked, what each check shows and what was not done. Five conditions matter for a \DSMC{} result of this kind: a time step short compared with the collision time, cells small compared with the mean free path, enough particles and enough sampling time, a flow that is stationary while it is sampled, and a result that does not depend on the grid. The first four are measured on the production runs. The fifth cannot be tested by grid refinement in three dimensions, and Section~\ref{app:grid} says what we use instead.

\subsection{Run ledger}
Table~\ref{tab:app_ledger} lists the production runs of the planar campaign and the three-dimensional runs of C1. The planar jet-on run of C1 took $4.9\times10^{3}$ core-hours on 1024 tasks, against $1.79\times10^{5}$ core-hours on 7680 tasks for the three-dimensional run, a factor of 37. A pair of jet-on and jet-off runs is needed for each $\cpi$, so the saving applies to both. The table lists the C1 pair as the representative three-dimensional run. The three-dimensional C2 and C3 runs use the same base grid ($175\times101\times46$ cells) and maximum refinement level (4), with the particle weight scaled with the freestream density.

\begin{table}[htbp]
  \centering
  \caption{Production runs, from the \SPARTA{} logs. Steps and wall time are summed over all phases, and core-hours are tasks times loop time. The particle count is the largest reached.}
  \label{tab:app_ledger}
  \small
  \begin{tabular}{@{}llllrrrr@{}}
    \toprule
    Case & Model & Run & Tasks & Steps & Wall [h] & Core-hours & Particles \\
    \midrule
    C1 & planar & jet-on & 1{,}024 & 438{,}202 & 4.8 & 4{,}878 & \num{1.09e+08} \\
    C2 & planar & jet-off & 1{,}024 & 211{,}501 & 1.6 & 1{,}638 & \num{1.40e+07} \\
    C2 & planar & jet-on & 1{,}024 & 502{,}602 & 4.7 & 4{,}840 & \num{9.53e+07} \\
    C3 & planar & jet-off & 1{,}024 & 211{,}301 & 1.5 & 1{,}542 & \num{1.39e+07} \\
    C3 & planar & jet-on & 512 & 276{,}701 & 1.6 & 839 & \num{1.56e+07} \\
    C1 & 3D & jet-off & 7{,}680 & 190{,}705 & 22.9 & 175{,}587 & \num{3.01e+09} \\
    C1 & 3D & jet-on & 7{,}680 & 190{,}705 & 23.3 & 178{,}686 & \num{3.09e+09} \\
    \bottomrule
  \end{tabular}
\end{table}

\subsection{Time step}
The move--collide splitting requires $\Delta t$ to be short compared with the collision time. The production time step is $\Delta t=\SI{1e-8}{s}$. The freestream collision times are 3.61, 2.07 and $\SI{0.73}{\micro s}$ for C1 to C3, so $\Delta t/\tau_\text{coll}=0.003$, 0.005 and 0.014. The densest gas is at the nozzle exit, where $\tau_\text{coll}=\SI{0.37}{\micro s}$ and the ratio is 0.027. The largest value in any cell of the planar C1 jet-off field is 0.010 (Table~\ref{tab:app_audit}). The runs also report the collisions per particle per step, averaged over the production phase, which is 0.006 to 0.026 (Table~\ref{tab:app_stat}), so at most about 3\% of the particles collide in a step. The Courant number $\Delta t\,u/\Delta$ is 0.15 in the planar freestream and 0.21 at the jet exit for the $\SI{80.5}{\micro m}$ finest cell, and 0.10 and 0.13 for the $\SI{125}{\micro m}$ finest cell of the three-dimensional grid, so a particle needs at least five steps to cross the smallest cell.

\subsection{Cell size}
The cell size is compared with the local mean free path $\lambda$ from the \SPARTA{} field itself. Table~\ref{tab:app_audit} gives the planar C1 jet-off field. In the first row of cells above the plate, which carries the wall pressure, the median $h/\lambda$ is 0.035 and no cell reaches $\lambda/3$. In the shock layer, 3.1\% of the area has cells larger than $\lambda/3$ and 0.3\% larger than $\lambda$. The undisturbed freestream, where the refinement criteria deliberately leave the coarse cells, has $h/\lambda=1.2$, which is harmless in a uniform stream. In the three-dimensional runs, $\lambda/\bar\Delta$ is evaluated in situ on every production dump, and in the wall row, which carries the pressure and every integrated force, $\Delta/\lambda=0.25$, 0.36 and 1.04 for C1 to C3.

The resolution is therefore ample for C1 and tighter for the denser cases, because the same finest cell is used while the mean free path falls by a factor of six. In the planar runs the finest cell is $0.15$, $0.31$ and $0.89$ of the freestream mean free path for C1 to C3, and the resolution audit of C3 puts its post-shock region at $L4/\lambda\approx4.2$, against the usual target of $\lambda/3$. We accepted this compromise for C3, as the three-dimensional runs accept coarser cells in the shock interior. It limits what we claim from C3 to the wall pressure and the interference lobe, which the near-wall cells carry, and not the internal structure of its shocks.

\begin{table}[htbp]
  \centering
  \caption{Resolution audit of the planar C1 jet-off field at step 132{,}800, from the cell bounds and the mean free path in the dump. $h$ is the larger cell side. The area fractions weight each cell by its area, and PPC is the sampled particle count per cell. The region definitions follow the text, and the undisturbed freestream is the region outside the refinement zone.}
  \label{tab:app_audit}
  \small
  \begin{tabular}{@{}lrrrrrr@{}}
    \toprule
    Region & Cells & Median $h/\lambda$ & Area $h>\lambda/3$ [\%] & Area $h>\lambda$ [\%] & Median PPC & Max $\Delta t/\tau$ \\
    \midrule
    First cell row above the plate & 4{,}531 & 0.035 & 0.0 & 0.0 & 0.33 & 0.0059 \\
    Plate region{,} $z<\SI{30}{mm}$ & 1{,}536{,}513 & 0.037 & 11.5 & 10.7 & 0.35 & 0.0101 \\
    Shock layer ($T>3T_\infty$) & 1{,}798{,}255 & 0.041 & 3.1 & 0.3 & 0.38 & 0.0088 \\
    Undisturbed freestream ($T<1.1T_\infty$) & 227{,}076 & 1.202 & 100.0 & 100.0 & 49.96 & 0.0031 \\
    \bottomrule
  \end{tabular}
\end{table}

\subsection{Particle weights and particles per cell}
Both campaigns use a particle weight chosen so that the finest cells next to the wall hold about one particle. The planar jet-on runs of C1 and C2 reduce the weight by a factor of 5.46 before production (Table~\ref{tab:2d_repro}), which gives about 4.4 particles in a finest freestream cell and 26 behind the shock. The planar jet-off runs of all three cases, and C3 in both modes, are sampled at the marking weight. That leaves about 0.8 particles in a finest freestream cell, and the median in the plate region of the C1 jet-off field is 0.35 (Table~\ref{tab:app_audit}). A particle count below one per cell is possible here because the cells are much smaller than the mean free path, so the pairs that collide are close neighbours, closer than the physical collision geometry requires. The cost is statistical noise in single-cell quantities, which time averaging over the production windows reduces. In the three-dimensional runs the jet-on and jet-off runs of C1 use the same weight of $4.77\times10^{11}$, and the 3.1$\times10^9$ particles in $1.1\times10^8$ cells give 28 particles per cell on average.

\subsection{Steady state and sampling}
Each case settles for three flow-through times before sampling in the jet-on runs and for 1.4 in the planar jet-off runs. Two production windows of 1.86 flow-through times each are then sampled, except in the asymmetric three-dimensional C2 runs, which sample one. Table~\ref{tab:app_stat} shows that the particle number is stationary through production. Its relative standard deviation is below 0.04\% in every run, and the drift from the first to the last quarter of production is below 0.05\%. The collision acceptance ratio, between 0.26 and 0.43, shows that the majorant of the collision scheme is not wasteful. The two production windows of the planar C1 runs also agree. The area-averaged pressure on the plate upstream of the slot changes between the windows by $-0.2$\% and 0.0\% in the jet-on runs of the coarse and medium grids, by $+0.05$\% in the production jet-on run, and by $-0.4$\%, $-0.1$\% and $-1.0$\% in the jet-off runs of the coarse, medium and production grids. On the plate downstream of the slot the changes are at most 3.6\%. The jet-off changes all have the same sign, so the baseline is still relaxing slowly after its short settle. We take this drift as the sampling uncertainty of the baseline. It corresponds to at most 0.002 in $\cpi$, against a lobe peak of 0.39.

The three-dimensional C1 runs show the same stationarity. Along the nozzle axis ($|y|<\SI{2.5}{mm}$, 2\,mm bins) the two windows give a lobe peak of 0.0335 and 0.0337 at $x=\SI{-7}{mm}$ and a downstream well of $-0.0306$ and $-0.0305$, so the half-difference of the windows is $10^{-4}$, or 0.3\% of the peak. Far upstream, at $x<\SI{-35}{mm}$, the rms half-difference is $8\times10^{-5}$, so the statements that the interference there is below $10^{-3}$ lie well above the noise floor. The jet-off pressure along the axis changes between the windows by 0.17\% on average, with no systematic sign. The result depends more on the choice of strip than on the sampling. Widening the half-width to $\SI{5}{mm}$ lowers the peak by 2.5\% and changes the well from $-0.031$ to $-0.022$, because the well is narrow. The values of Table~\ref{tab:cpi_metrics} are read from the plotted curves and agree with the narrow strip to within 5\% (peak 0.032 against 0.034, well $-0.030$ against $-0.031$), with the peak position one bin away.

\begin{table}[htbp]
  \centering
  \caption{Stationarity of the production phase, from the \SPARTA{} logs. $N_p$ is the particle number sampled at each logged step. The drift compares the mean over the last quarter of the phase with the first. The collision rate is $2N_\text{coll}/N_p$ per step, and the acceptance ratio is accepted over attempted collisions.}
  \label{tab:app_stat}
  \small
  \begin{tabular}{@{}lllrrrrr@{}}
    \toprule
    Case & Model & Run & Steps & $N_p$ std [\%] & $N_p$ drift [\%] & Collisions per particle per step & Accept ratio \\
    \midrule
    C1 & planar & jet-on & 132{,}800 & 0.012 & +0.017 & 0.0109 & 0.36 \\
    C2 & planar & jet-off & 153{,}200 & 0.023 & -0.028 & 0.0059 & 0.43 \\
    C2 & planar & jet-on & 153{,}200 & 0.013 & +0.019 & 0.0146 & 0.35 \\
    C3 & planar & jet-off & 153{,}000 & 0.022 & -0.032 & 0.0159 & 0.42 \\
    C3 & planar & jet-on & 152{,}800 & 0.018 & -0.022 & 0.0259 & 0.31 \\
    C1 & 3D & jet-off & 104{,}904 & 0.035 & -0.011 & 0.0061 & 0.39 \\
    C1 & 3D & jet-on & 104{,}904 & 0.033 & -0.047 & 0.0065 & 0.26 \\
    \bottomrule
  \end{tabular}
\end{table}

\subsection{Dependence on the grid}
\label{app:grid}
The planar C1 case was run on three grids (Table~\ref{tab:app_levels}). The base grids of $86\times50$, $172\times100$ and $688\times400$ cells are refined with thresholds scaled to the base cell, which gives finest cells of 643, 321 and $\SI{80}{\micro m}$, so the cell size changes by factors of two and four. The time step falls from $10^{-7}$ to $5\times10^{-8}$ and $10^{-8}\,\text{s}$, and the particle weight scales with the cell area, so the median particle count in the wall row stays at 0.35 in the jet-off runs of all three levels. The sweep is a joint refinement of cell size and time step with the statistical quality held fixed. All three levels are evaluated in the same way, from the wall-row cell pressure of the final grid dumps in 2\,mm bins, except for the area averages, which come from the surface dumps.

\begin{table}[htbp]
  \centering
  \caption{Resolution sweep of the planar C1 case. The wall-row $h/\lambda$ is the median of the jet-off run, $\langle p\rangle$ is the area-averaged pressure on the plate upstream of the slot in the jet-off run, and $\langle\Delta p\rangle$ is its increase when the jet is on. The interference values are the lobe peak and the value at $x=\SI{-40}{mm}$. The order and the limit follow from a three-level Richardson analysis of each column.}
  \label{tab:app_levels}
  \small
  \begin{tabular}{@{}lrrrrrrr@{}}
    \toprule
    Grid & Finest cell [\si{\micro m}] & $\Delta t$ [s] & $h/\lambda$ & $\langle p\rangle$ [Pa] & $\langle\Delta p\rangle$ [Pa] & $\cpi$ peak & $\cpi(-40)$ \\
    \midrule
    Coarse & 643 & $10^{-7}$ & 0.29 & 3.579 & 8.98 & 0.379 & 0.286 \\
    Medium & 321 & $5\times10^{-8}$ & 0.15 & 3.547 & 9.90 & 0.386 & 0.305 \\
    Production & 80 & $10^{-8}$ & 0.036 & 3.525 & 10.37 & 0.391 & 0.310 \\
    \midrule
    Observed order & & & & 1.1 & 1.4 & 1.1 & 2.0 \\
    Extrapolated limit & & & & 3.519 & 10.44 & 0.392 & 0.311 \\
    \bottomrule
  \end{tabular}
\end{table}

\begin{figure}[htbp]
  \centering
  \includegraphics[width=\linewidth]{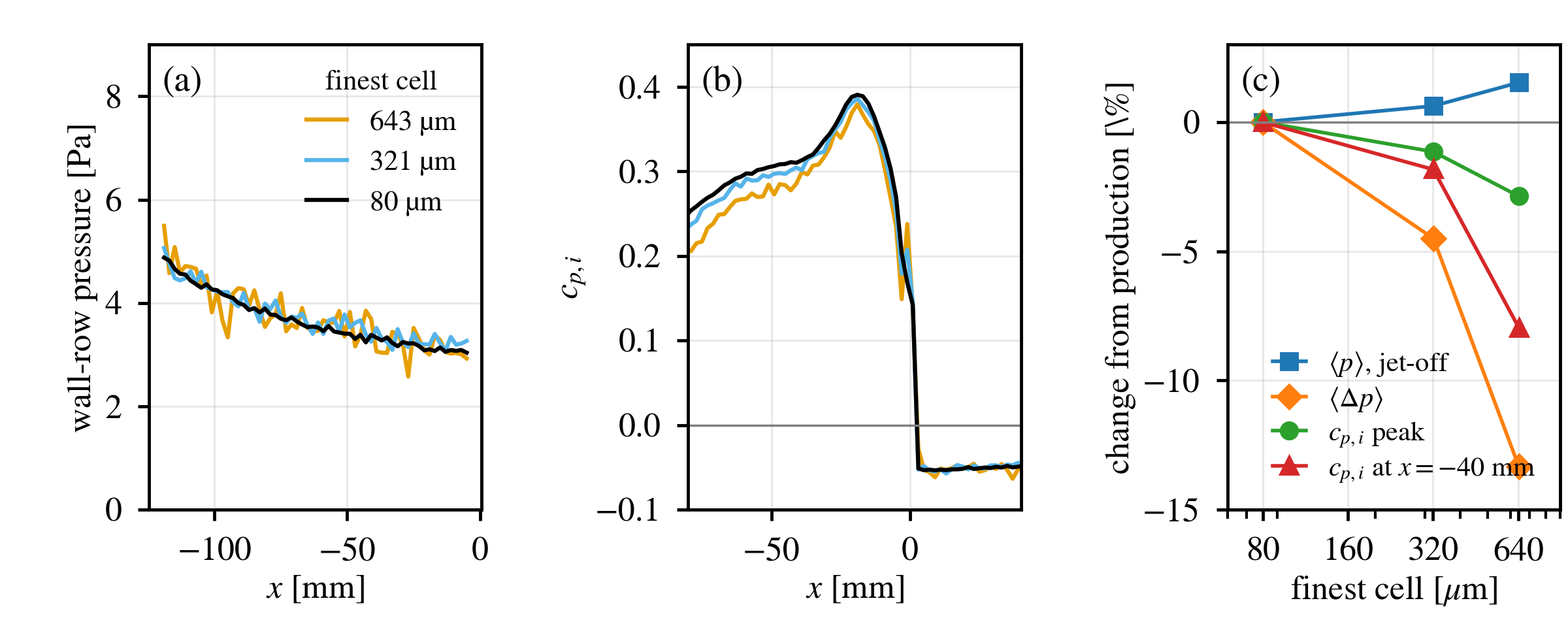}
  \caption{Resolution sweep of the planar C1 case. (a) Jet-off wall-row pressure on the three grids. (b) Interference pressure coefficient on the coarse (orange), medium (blue) and production (black) grids. (c) Change from the production value against the finest cell size, for the jet-off plate pressure, the area-averaged interference $\langle\Delta p\rangle$, and the interference pressure at its peak and at $x=\SI{-40}{mm}$.}
  \label{fig:app_levels}
\end{figure}

The jet-off wall pressure converges quickly. The plate average is 3.579, 3.547 and \SI{3.525}{Pa}, a change of 0.9\% and then 0.6\%. Bin by bin over $x=-120$ to $\SI{-6}{mm}$ the mean differences are 6.9\% between the coarse and medium grids and 3.6\% between the medium and production grids, with maxima of 21\% and 11\% at $x=-95$ and $\SI{-51}{mm}$. The ratio of the two successive differences is 2.0, between the 1.3 expected for first-order convergence and the 3.2 for second order, for these refinement ratios. For the plate average the observed order is 1.1, and Richardson extrapolation gives 3.519 Pa, within 0.2\% of the production value.

The interference converges too, in every measure we examined. The area-averaged interference pressure on the plate upstream of the slot, the jet-on average minus the jet-off average, is 8.98, 9.90 and \SI{10.37}{Pa}, a change of 10.2\% and then 4.7\%. The ratio of the successive differences, 2.0, gives an observed order of 1.4 and a Richardson limit of \SI{10.44}{Pa}, so the production value is within 0.7\% of it, while the medium and coarse grids lie 5\% and 14\% below. The lobe peak is 0.379, 0.386 and 0.391, a change of 1.8\% and then 1.2\%, with an observed order of 1.1 and a limit of 0.392, so the production peak is within 0.3\% of it. At $x=\SI{-40}{mm}$, where the planar lobe is still 80\% of its peak, the values are 0.286, 0.305 and 0.310, with an observed order of 2.0 and a limit of 0.311, and the production value is within 0.1\% of it. The integral of $\cpi$ over $x=-60$ to $\SI{-3}{mm}$ is 16.7, 17.5 and \SI{18.0}{mm}, a change of 4.6\% and then 2.6\%, with an order of 1.3 and a limit of 18.1, and the production value is within 0.5\%. Table~\ref{tab:cpi_metrics}, which is read from the unbinned production curve, gives 0.394 and 0.315 for the first two quantities, which are 0.9\% and 1.5\% higher than the binned values used here. The planar lobe at $\SI{40}{mm}$ ahead of the nozzle is above 0.28 on every grid, against less than $10^{-3}$ in three dimensions, so the difference in upstream extent does not depend on the grid.

We could run this sweep only in two dimensions and only for C1. C1 stands in for C2 and C3 because the three cases use the same grid ($688\times400$ base cells, the same refinement criteria and the same $\SI{80}{\micro m}$ finest cell), the same time step and the same collision model, and differ only in the freestream and jet conditions (Table~\ref{tab:cases}). C2 and C3 are less well resolved than C1 relative to the mean free path, and C3 least, with a finest cell of 0.89 of the freestream mean free path and $L4/\lambda\approx4.2$ behind the shock. The sweep therefore shows the numerical procedure to converge where it is well resolved, and it does not by itself demonstrate convergence of C3, for which we claim only the wall pressure and the interference lobe. A refinement of the three-dimensional grid multiplies the cell count by about eight per level, from $1.1\times10^8$ cells in the asymmetric C1 run, and that run already took $1.8\times10^5$ core-hours. The three-dimensional runs use the same collision model, wall condition and refinement criteria, and their finest cells are comparable to the planar production cells relative to the mean free path ($\SI{125}{\micro m}$ gives 0.23 of the freestream value for C1, against 0.15). The planar sweep therefore stands in for them, and it says nothing independent about the three-dimensional quantities. For those, the check is the agreement of the lobe with Hefer's measurements to within 25\%.

Two further observations point the same way. First, the planar overprediction is the same at three densities whose resolution differs by a factor of six. The finest planar cell is 0.15, 0.31 and 0.89 of the freestream mean free path for C1 to C3, and the peak interference pressure is 11.7, 11.5 and 11.8 times Hefer's, constant to within 3\% (Table~\ref{tab:cpi_metrics}). Second, the three-dimensional reference has finest cells that are coarser than the planar ones relative to the freestream mean free path (0.23, 0.48 and 1.4 against 0.15, 0.31 and 0.89), yet it reproduces Hefer's lobe. Together with the sweep, these bound the effect of the grid on the central result: the planar lobe peak, its value at 40 mm ahead of the nozzle, its integral and the area-averaged interference are each within 0.1\% to 0.7\% of their extrapolated limits on the production grid. They do not amount to a convergence study of every quantity, and we do not claim one.

\end{appendix}

\begin{acknowledgements}
The authors thank the DLR Institute of Aerodynamics and Flow Technology for
access to the CARO cluster.
\end{acknowledgements}

\section*{Statements and Declarations}
\noindent\textbf{Funding.} The work was funded by DLR G\"ottingen.\\
\textbf{Competing interests.} The authors have no relevant financial or non-financial interests to disclose.\\
\textbf{Data availability.} The digitised Hefer data and the simulation data will be
provided by the first author (A.T.) on request.\\
\textbf{Code availability.} \SPARTA{} is open-source software \citep{sparta}; the
input decks and the post-processing scripts will be provided by the first author (A.T.) on request.\\
\textbf{Author contributions.} A.T.: conceptualisation, methodology, software,
validation, formal analysis, investigation, data curation, visualisation,
writing (original draft). M.G.: conceptualisation, supervision, resources.

\bibliographystyle{spmpsci_cit}
\bibliography{refs}

\begin{thebibliography}{10}
\providecommand{\url}[1]{{#1}}
\providecommand{\urlprefix}{URL }
\expandafter\ifx\csname urlstyle\endcsname\relax
  \providecommand{\doi}[1]{DOI~\discretionary{}{}{}#1}\else
  \providecommand{\doi}{DOI~\discretionary{}{}{}\begingroup
  \urlstyle{rm}\Url}\fi

\bibitem{spaid1973}
Spaid, F.W., Cassel, L.A.: Aerodynamic interference induced by reaction
  controls.
\newblock AGARDograph AGARD-AG-173, AGARD, Neuilly-sur-Seine, France (1973)

\bibitem{hefer1975}
Hefer, G.: {Experimentelle Untersuchungen zur Interferenz zwischen einer
  Hyperschallstr{\"o}mung geringer Dichte und einem Strahlspoiler}.
\newblock Ph.D. thesis, Technische Universit\"{a}t Braunschweig (1975).
\newblock DFVLR-AVA Bericht Nr.\ IB~252-75~H~01

\bibitem{hefer1976}
Hefer, G.: {Experimentelle Untersuchungen zur Interferenz zwischen einer
  Hyperschallstr{\"o}mung geringer Dichte und einem Strahlspoiler}.
\newblock Forschung im Ingenieurwesen \textbf{42}(4), 123--130 (1976).
\newblock \doi{10.1007/BF02573854}

\bibitem{powrie1993}
Powrie, H.E.G., Ball, G.J., East, R.A.: Comparison of the interactions of two
  and three dimensional transverse jets with hypersonic free stream.
\newblock In: Computational and Experimental Assessment of Jets in Cross Flow,
  AGARD-CP-534. Winchester, UK (1993).
\newblock Paper 20, pp.~20-1--20-8

\bibitem{powrie1996}
Powrie, H.E.G.: A study of the interaction between an underexpanded normal jet
  and a hypersonic free stream.
\newblock Ph.D. thesis, University of Southampton (1996).
\newblock 359 pp.

\bibitem{dixon2005}
Dixon, F.C., Roberts, G.T., Tutty, O.R.: Interference effects due to a
  transverse slot jet in a hypersonic flow.
\newblock In: Z.~Jiang (ed.) Shock Waves, pp. 137--142. Springer, Berlin,
  Heidelberg (2005).
\newblock \doi{10.1007/978-3-540-27009-6\_17}

\bibitem{mahesh2013interaction}
Mahesh, K.: The interaction of jets with crossflow.
\newblock Annual Review of Fluid Mechanics \textbf{45}, 379--407 (2013).
\newblock \doi{10.1146/annurev-fluid-120710-101115}

\bibitem{karagozian2014}
Karagozian, A.R.: The jet in crossflow.
\newblock Physics of Fluids \textbf{26}(10), 101303 (2014).
\newblock \doi{10.1063/1.4895900}

\bibitem{Chenhao2014}
Xu, C., Jiang, C., Gao, Z., Qian, Z., Li, C.: The jet interaction effects of
  reaction control systems in hypersonic vehicles.
\newblock Mechanics in Engineering \textbf{36}(2) (2014).
\newblock \doi{10.6052/1000-0879-13-376}.
\newblock In Chinese, with English abstract

\bibitem{tartabini1995}
Tartabini, P.V., Wilmoth, R.G., Rault, D.F.G.: Direct simulation {Monte Carlo}
  calculation of a jet interaction experiment.
\newblock Journal of Spacecraft and Rockets \textbf{32}(1), 75--83 (1995).
\newblock \doi{10.2514/3.26577}

\bibitem{Dettleff1991}
Dettleff, G.: Plume flow and plume impingement in space technology.
\newblock Progress in Aerospace Sciences \textbf{28}(1), 1--71 (1991).
\newblock \doi{10.1016/0376-0421(91)90008-R}

\bibitem{Yaryginetal2011}
Yarygin, V.N., Gerasimov, Y.I., Krylov, A.N., Mishina, L.V., Prikhodko, V.G.,
  Yarygin, I.V.: Gas dynamics of spacecraft and orbital stations (review).
\newblock Thermophysics and Aeromechanics \textbf{18}(3), 333--358 (2011).
\newblock \doi{10.1134/s0869864311030012}

\bibitem{vogenitz1970}
Vogenitz, F.W., Broadwell, J.E., Bird, G.A.: Leading edge flow by the {M}onte
  {C}arlo direct simulation technique.
\newblock AIAA Journal \textbf{8}(3), 504--510 (1970).
\newblock \doi{10.2514/3.5699}

\bibitem{Vogenitzetal1971}
Vogenitz, F.W., Takata, G.Y.: Rarefied hypersonic flow about cones and flat
  plates by {M}onte {C}arlo simulation.
\newblock AIAA Journal \textbf{9}, 94--100 (1971).
\newblock \doi{10.2514/3.6128}

\bibitem{ivanov1998}
Ivanov, M.S., Gimelshein, S.F.: Computational hypersonic rarefied flows.
\newblock Annual Review of Fluid Mechanics \textbf{30}, 469--505 (1998).
\newblock \doi{10.1146/annurev.fluid.30.1.469}

\bibitem{ivanov1998aiaa}
Ivanov, M.S., Markelov, G.N., Gimelshein, S.F.: Statistical simulation of
  reactive rarefied flows: numerical approach and applications.
\newblock AIAA Paper 98-2669 (1998)

\bibitem{tartabini1993}
Tartabini, P.V., Wilmoth, R.G., Rault, D.F.G.: A systems approach to a {DSMC}
  calculation of a control jet interaction experiment.
\newblock AIAA Paper 93-2798, 28th Thermophysics Conference, Orlando, FL (1993)

\bibitem{Glass2018}
Glass, C.E.: Numerically simulating an expanding continuum jet into a
  surrounding non-continuum region.
\newblock Tech. Rep. NASA/TP-2018-220095, NASA Langley Research Center (2018)

\bibitem{glass1997}
Glass, C.E., LeBeau, G.J.: Numerical study of a continuum sonic jet interacting
  with a rarefied flow.
\newblock AIAA Paper 97-2536 (1997)

\bibitem{liang2014}
Liang, J., Li, Z., Li, X., Fang, M.: {DSMC} numerical simulation of lateral jet
  interaction with rarefied atmosphere.
\newblock In: Proceedings of the 29th International Symposium on Rarefied Gas
  Dynamics, vol. 1628, pp. 589--595. AIP (2014).
\newblock \doi{10.1063/1.4902646}

\bibitem{Glass2000}
Glass, C.E.: Numerical study of rarefied hypersonic flow interacting with a
  continuum jet.
\newblock Technical Publication NASA/TP-2000-210551, NASA Langley Research
  Center (2000)

\bibitem{glass2004}
Glass, C.E.: A parametric study of jet interactions with rarefied flow.
\newblock Tech. Rep. NTRS 20040095945, NASA Langley Research Center (2004)

\bibitem{Staack2001}
Staack, D., McDaniel, J.C., Glass, C.E., Miller, C.: Experimental study of
  interacting rarefied and continuum flows.
\newblock In: 35th AIAA Thermophysics Conference. Anaheim, CA (2001).
\newblock \doi{10.2514/6.2001-2762}.
\newblock AIAA Paper 2001-2762

\bibitem{gimelshein2002}
Gimelshein, S.F., Alexeenko, A.A., Levin, D.A.: Modeling of the interaction of
  a side jet with a rarefied atmosphere.
\newblock Journal of Spacecraft and Rockets \textbf{39}(2), 168--176 (2002).
\newblock \doi{10.2514/2.3814}

\bibitem{kashkovsky2014}
Kashkovsky, A.V., Vashchenkov, P.V., Banyai, T.: Modeling of reentry space
  vehicle aerodynamics with control thruster plume --- free-stream interaction.
\newblock Thermophysics and Aeromechanics \textbf{21}(6), 719--728 (2014).
\newblock \doi{10.1134/S0869864314060067}

\bibitem{Karpuzcu2023}
Karpuzcu, I.T., Levin, D.A.: Study of side-jet interactions over a hypersonic
  cone flow using kinetic methods.
\newblock AIAA Journal \textbf{61}(11), 4741--4751 (2023).
\newblock \doi{10.2514/1.j063020}

\bibitem{sparta}
Plimpton, S.J., Moore, S.G., Borner, A., Stagg, A.K., Koehler, T.P.,
  Torczynski, J.R., Gallis, M.A.: Direct simulation {Monte Carlo} on petaflop
  supercomputers and beyond.
\newblock Physics of Fluids \textbf{31}(8), 086101 (2019).
\newblock \doi{10.1063/1.5108534}

\bibitem{theofilis2026companion}
Theofilis, A., Grabe, M.: Manuscript in preparation (2026)

\bibitem{theofilis2026icjwsf}
Theofilis, A., Grabe, M.: {Numerical Study of the Interaction of a Jet
  Exhausting Into Rarefied Hypersonic Crossflow}.
\newblock Presentation at the 9th International Conference on Jets, Wakes and
  Separated Flows (ICJWSF 2026), Tucson, AZ, USA, 20--22 October (2026)

\bibitem{theofilis2026dfd}
Theofilis, A., Grabe, M.: {Numerical Study of the Interaction of a Jet
  Exhausting Into Rarefied Hypersonic Crossflow}.
\newblock Presentation at the 79th Annual Meeting of the APS Division of Fluid
  Dynamics, Orlando, FL, USA, 22--24 November (2026)

\bibitem{bird1994}
Bird, G.A.: Molecular Gas Dynamics and the Direct Simulation of Gas Flows.
\newblock Oxford University Press, Oxford (1994)

\bibitem{Klothakis2022}
Klothakis, A., Quintanilha, H., Sawant, S.S., Protopapadakis, E., Theofilis,
  V., Levin, D.A.: Linear stability analysis of hypersonic boundary layers
  computed by a kinetic approach: a semi-infinite flat plate at {$4.5 \le
  M_\infty \le 9$}.
\newblock Theoretical and Computational Fluid Dynamics \textbf{36}(1), 117--139
  (2022).
\newblock \doi{10.1007/s00162-021-00601-y}

\bibitem{klothakis2021aiaa}
Klothakis, A., Sawant, S.S., Quintanilha~Jr., H., Theofilis, V., Levin, D.A.:
  Slip effects on the stability of supersonic laminar flat plate boundary
  layer.
\newblock In: AIAA Scitech 2021 Forum (2021).
\newblock \doi{10.2514/6.2021-1659}.
\newblock AIAA 2021-1659

\end{thebibliography}

\end{document}